\documentclass[a4paper,11pt]{article}
\pdfoutput=1 

\usepackage{jcappub} 

\usepackage[T1]{fontenc} 
\usepackage{ifthen}
\usepackage{epsfig}
\usepackage{booktabs} 
\usepackage{natbib}
\usepackage{bbold}
\usepackage{diagbox}
\usepackage{gensymb}

\newcommand{\erf}{\mathop{\mathrm{erf}}}
\newcommand{\beqra}{\begin{flalign}}
\newcommand{\eeqra}{\end{flalign}}
\newcommand{\beq}{\begin{equation}}
\newcommand{\eeq}{\end{equation}}

\newcommand{\OmCap}{\Omega_v^-(w)}

\usepackage{ccicons} 
\usepackage{listings}
\usepackage{xcolor}   
\usepackage{braket}  
\usepackage{hyperref}
\usepackage{dsfont}  
\usepackage{leftidx} 
\RequirePackage{mathrsfs}
\RequirePackage{dsfont}
\usepackage{bm}
\usepackage{placeins}
\usepackage{float}
\allowdisplaybreaks

\title{\boldmath Assessing the sensitivity of PINGU to effective dark matter-nucleon interactions}
\author[a]{Anton B\"ackstr\"om}
\author[a]{\hspace{-0.25 cm}, Riccardo Catena}
\author[b]{and Carlos P\'erez de los Heros}
\affiliation[a]{Chalmers University of Technology, Department of Physics, SE-412 96 G\"oteborg, Sweden}
\affiliation[b]{Department of Physics and Astronomy, Uppsala University, Box 516, SE-751 20 Uppsala, Sweden}
\emailAdd{antonba@student.chalmers.se}
\emailAdd{catena@chalmers.se}
\emailAdd{cph@physics.uu.se}

\abstract{We calculate the sensitivity of next generation neutrino telescopes to the 28 (isoscalar and isovector) coupling constants defining the non-relativistic effective theory of (spin 1/2) dark matter (DM)-nucleon interactions.~We take as a benchmark detector the proposed Precision IceCube Next Generation Upgrade (PINGU), although our results are valid for any other neutrino telescope of similar effective volume.~We express PINGU's sensitivity in terms of $5\sigma$ sensitivity contours in the DM-mass -- coupling constant plane, and compare our sensitivity contours with the 90\% C.L. exclusion limits on the same coupling constants that we obtain from a reanalysis of the null result of current DM searches at IceCube/DeepCore.~We find that PINGU can effectively probe not only the canonical spin-independent and spin-dependent DM-nucleon interactions, but also velocity-dependent or momentum-dependent interactions that generate coherently enhanced DM-nucleus scattering cross sections.~We also find that PINGU's $5\sigma$ sensitivity contours are significantly below current IceCube/DeepCore 90\% C.L. exclusion limits when $b\bar{b}$ is the leading DM annihilation channel.~This result shows the importance of lowering the experimental energy threshold when probing models that generate soft neutrino energy spectra, and holds true independently of the assumed DM-nucleon interaction and for all DM masses tested here.~When DM primarily annihilates into $\tau\bar{\tau}$, a PINGU-like detector will improve upon current exclusion limits for DM masses below $35$~GeV, independently of the assumed DM-nucleon interaction.}

\begin{document}
\maketitle
\flushbottom

\section{Introduction}
Elucidating the nature of dark matter (DM) remains one of the major unsolved questions in modern physics~\cite{Bertone:2016nfn}.~In the leading paradigm, DM is made of weakly interacting massive particles, WIMPs.~WIMPs are hypothetical particles with interactions at the weak scale, typical masses in the 1 GeV -- 100 TeV range, and present cosmological density set in the early Universe by the chemical decoupling mechanism~\cite{Arcadi:2017kky}.~Because of these properties, existing experimental techniques can be used to search for non gravitational signals of WIMP DM~\cite{Roszkowski:2017nbc}.~These include direct detection experiments searching for DM-nucleus or -electron interactions in low-background environments~\cite{Undagoitia:2015gya}, indirect detection experiments searching for DM annihilation products with space satellites~\cite{Gaskins:2016cha},  ground-based observatories~\cite{Doro:2014pga} or neutrino telescopes~\cite{Achterberg:2006md,Fukuda:2002uc,Collaboration:2011nsa,Avrorin:2011zza,Alekseev:1998ib}, and particle accelerators searching for DM as missing transverse momentum in the final state of energetic collisions between Standard Model particles~\cite{Vartak:2017yfz,Kalderon:2018hdv}.~A unique feature of neutrino telescopes is the ability to simultaneously probe the physics of DM annihilation (as indirect detection experiments) and the nature of DM-nucleus scattering (similarly to direct detection experiments).~This is a consequence of the physical process neutrino telescopes aim at probing~\cite{losHeros:2017otx}.~They search for neutrinos from the annihilation of DM particles which in the past few billion years accumulated at the centre of the Sun (or of the Earth) by losing energy via DM-nucleus scattering while crossing the solar (or terrestrial) interior~\cite{Krauss:1985aaa,Krauss:1985ks,Freese:1985qw,PhysRevD.39.1029}.

The theoretical modelling of DM capture and annihilation in the Sun (or in the Earth) involves several non trivial steps~\cite{,Gould:1987ww,Gould:1987ir,PhysRevD.39.1029,Gould:1991rc,Gould:1999je,Blennow:2007tw,Wikstrom:2009kw,Sivertsson:2012qj,Blennow:2015hzp,Blennow:2018xwu,Catena:2018vzc}.~Firstly, one has to calculate the rate at which DM particles are captured in the Sun.~The rate of DM capture in the Sun depends on the cross section for DM-nucleus scattering (which requires particle and nuclear physics inputs), on the kinematics of DM-nucleus scattering (elastics vs.~inelastic scattering), on the chemical composition of the Sun, and, finally, on the velocity distribution of DM particles in the solar neighbourhood.

Secondly, one has to calculate the rate at which DM particles pair annihilate in the Sun.~The DM pair annihilation rate depends on whether or not dynamical equilibrium between capture and annihilation has been reached during the Sun's lifetime.~If not, it explicitly depends on the thermally averaged DM annihilation cross section (times relative velocity of the annihilating DM particles) and on the spatial distribution of DM in the Sun, which is not necessarily isothermal.

Finally, the DM pair annihilation rate has to be converted into the measurable flux of neutrinos at the detector.~The neutrino flux from DM annihilation in the Sun depends on the branching ratios for DM pair annihilation into Standard Model final states, on the energy spectra of neutrinos produced from the decay of the particles in the allowed final states, on the transport of neutrinos from where they are produced to the detector, and on a number of experimental inputs, e.g.~angular resolution and energy threshold (which is encoded in the detector effective area).

From the experimental side, neutrino telescopes currently searching for neutrinos from DM annihilation in the Sun (or in the Earth) include IceCube, Super-Kamiokande, ANTARES, Baksan and Baikal~\cite{Achterberg:2006md,Fukuda:2002uc,Collaboration:2011nsa,Avrorin:2011zza,Alekseev:1998ib}.~While the experimental effort has been remarkable, DM has so far escaped detection at neutrino telescopes, as well as at experiments using different methods.~The null result of the present generation of neutrino telescopes has been used to set exclusion limits on the strength of DM-nucleon interactions, under different assumption regarding how DM pair annihilates~\cite{Aartsen:2016fep,Albert:2016dsy,Aartsen:2016zhm,Adrian-Martinez:2016gti,Choi:2015ara,Avrorin:2014swy,Boliev:2013ai}, and how DM interacts with nuclei (or electrons~\cite{Garani:2017jcj}) in the Sun~\cite{Catena:2015iea} (or in the Earth~\cite{Catena:2016kro,Baum:2016oow}).~Next generation neutrino telescopes will include:~1) The Baikal-GVD detector~\cite{Avrorin:2018ijk} in Lake Baikal, Russia, designed to detect astrophysical neutrinos at energies from a few TeV up to 100 PeV.~2) The KM3NeT research infrastructure~\cite{Adrian-Martinez:2016fdl} in the Mediterranean sea, housing the neutrino telescopes ARCA optimized for the search for high-energy  neutrinos from distant astrophysical sources (such as active galactic nuclei, gamma ray bursters or colliding stars), and a low-energy array, ORCA (which is optimised to study atmospheric neutrino oscillations); and 3) IceCube-Gen2~\cite{Aartsen:2014njl} at the South Pole, whose scientific goal is to record large samples of astrophysical neutrinos in the PeV to EeV range, and yield hundreds of neutrinos across all flavors at energies above 100 TeV.~IceCube-Gen-2 will also include a low-energy core with a denser sensor configuration which will target the determination of the neutrino mass hierarchy, precision measurements of the atmospheric oscillation parameters, and search for DM in a mass range which is currently unexplored at neutrino telescopes.

Motivated by the rapid progress the field of DM searches at neutrino telescopes has experienced in recent years, this article aims at a critical assessment of the sensitivity of next generation neutrino telescopes to DM-nucleon interactions.~We will address this question within the non-relativistic effective theory of DM-nucleon interactions~\cite{Fan:2010gt,Fitzpatrick:2012ix,Cirigliano:2012pq,Anand:2013yka,DelNobile:2013sia,Catena:2014uqa,Catena:2014epa,Catena:2014hla,Catena:2015vpa,Catena:2015uua,Hoferichter:2015ipa,Catena:2016hoj,Catena:2016ckl,Kavanagh:2016pyr,Catena:2017wzu,Baum:2017kfa,Catena:2017xqq,DelNobile:2018dfg,Kang:2018odb,Kang:2018qvz} (see also~\cite{Crivellin:2014qxa,DEramo:2014nmf,Bishara:2016hek,Brod:2017bsw,Bishara:2017pfq,Bishara:2017nnn,Bishara:2018vix,Kang:2018rad} for a discussion on relativistic effective theories for DM-quark and -gluon interactions).~We use the proposed Precision IceCube Next Generation Upgrade (PINGU)~\cite{LoI} as a benchmark detector for our studies, clarifying its physics reach and providing additional motivations to further investigate the phenomenology of low-mass WIMP DM with future similar neutrino telescopes.

This article is organised as follows.~In Sec.~\ref{sec:theory} we review the theory of DM capture and annihilation in the Sun.~Sec.~\ref{sec:analysis} is devoted to a description of how we calculate the sensitivity of PINGU to DM-nucleon interactions, while results are presented in Sec.~\ref{sec:results}.~We finally conclude in Sec.~\ref{sec:conclusion}.

\section{Theoretical framework}
\label{sec:theory}
In this section, we review the theory of DM capture in the Sun (Sec.~\ref{sec:capture}) and the predicted neutrino flux at neutrino telescopes (Sec.~\ref{sec:annihilation}).~This prediction depends on the cross section for DM-nucleus scattering, which we calculate within the effective theory of DM-nucleon interactions (Sec.~\ref{sec:eft}).

\subsection{Dark matter capture in the Sun}
\label{sec:capture}
For the WIMPs forming our galactic halo, the rate of scattering from an initial velocity $w$ to a velocity less than the local escape velocity at a distance $r$ from the Sun's centre, $v(r)$, is given by~\cite{Catena:2015uha}
\begin{align}
\Omega_{v}^{-}(w)= \sum_T n_T w\,\Theta\left( \frac{\mu_T}{\mu^2_{+,T}} - \frac{u^2}{w^2} \right)\int_{E u^2/w^2}^{E \mu_T/\mu_{+,T}^2} {\rm d}E_r\,\frac{{\rm d}\sigma_{\chi T}\left(E_r,w^2\right)}{{\rm d}E_r}\,,
\label{eq:omega}
\end{align}
where ${\rm d}\sigma_{\chi T}/{\rm d}E_r$ is the differential cross section for DM-nucleus scattering, $E_r$ the nuclear recoil energy deposited in the collision, $w=\sqrt{u^2+v^2(r)}$ the DM particle velocity in the target nucleus rest frame at a distance $r$ from the centre of the Sun, $u$ the speed such a particle would have at $r\rightarrow\infty$, $\mu_+=(\mu_T+1)/2$, where $\mu_T$ is the WIMP-nucleus reduced mass, $E=m_\chi w^2/2$, and the index $T$ runs over the 16 most abundant elements in the Sun, namely H, $^{3}$He, $^{4}$He, $^{12}$C, $^{14}$N, $^{16}$O, $^{20}$Ne, $^{23}$Na, $^{24}$Mg, $^{27}$Al, $^{28}$Si, $^{32}$S, $^{40}$Ar, $^{40}$Ca, $^{56}$Fe, and $^{58}$Ni.~The density of the $T$-th element at a distance $r$ from the Sun's centre is denoted by $n_T(r)$ and modelled as in the  {\texttt{DARKSUSY}} package, which also provides an expression for $v(r)$ in terms of the Sun's gravitational potential~\cite{Gondolo:2004sc}.~Multiplying Eq.~(\ref{eq:omega}) by the rate at which DM particles cross an infinitesimal solar shell at $r$ and the time spent by each DM particle on that shell, one obtains the differential rate at which WIMPs with velocity at infinity around $u$ are captured at a distance $r$ from the Sun's centre.~By integrating the latter over all radii $r$ and velocities $u$ that can contribute to WIMP capture, one finds for the total rate of DM capture in the Sun~\cite{Gould:1987ir}:
\begin{align}
C= \frac{\rho_\chi}{m_\chi} \int_0^{r_\odot} {\rm d}r \,4\pi r^2\int_0^\infty {\rm d}u \frac{f(u)}{u}w\OmCap \,,
\label{eq:C}
\end{align}
where $r_\odot$ is the solar radius, $m_\chi$ the DM particle mass, $\rho_\chi$ the density of DM particles in the solar neighbourhood, and $f(u)$ the DM speed distribution in the Sun's rest frame, for which we assume~\cite{Lewin:1995rx}
\begin{align}
f(u) &= \left\{
\begin{array}{ll}
\mathscr{N} 
\left[ \exp\left(-\frac{(u-v_{\rm obs})^2}{v_\odot^2}\right) - \exp\left(-\frac{(u+v_{\rm obs})^2}{v_\odot^2}\right)\right]& \quad \quad \text{for  } u\le (v_{\rm esc}-v_{\rm obs})\\
&\\
\mathscr{N} 
\left[ \exp\left(-\frac{(u-v_{\rm obs})^2}{v_\odot^2}\right) - \exp\left(-\frac{v^2_{\rm esc}}{v^2_{\odot}}\right)\right]&  \quad \quad \text{for  } (v_{\rm esc}-v_{\rm obs})<u\le (v_{\rm esc}+v_{\rm obs})\\
&\\
0 &  \quad \quad \text{for  } u> (v_{\rm esc}+v_{\rm obs}) \,,
\end{array}
\right. 
\end{align}
where
\begin{align}
\mathscr{N} &=\frac{1}{\pi^{3/2} v^3_{\rm obs}} 
\left[ \erf\left(\frac{v_{\rm esc}}{v_\odot}\right) -\frac{2}{\sqrt{\pi}} \left( \frac{v_{\rm esc}}{v_\odot}\right) \exp\left(-\frac{v^2_{\rm esc}}{v^2_\odot}\right) \right]^{-1} \,.
\end{align}
In our calculations, we set $\rho_\chi=0.4$~GeV/cm$^{3}$, $v_{\rm esc}=544$~km~s$^{-1}$, $v_\odot=220$~km~s$^{-1}$ and $v_{\rm obs}=232$~km~s$^{-1}$, e.g.~\cite{Freese:2012xd,Bozorgnia:2013pua,Catena:2011kv,Catena:2009mf}.

\subsection{Neutrinos from dark matter annihilation in the Sun}
\label{sec:annihilation}
The average number of WIMP annihilations per unit time in the Sun's interior, $\Gamma_a$, is given by
\begin{equation}
\Gamma_a=\frac{1}{2}\int d^3{\bf x} \,\epsilon^2({\bf x}) \,\langle \sigma_{\rm ann} v_{\rm rel} \rangle \,,
\label{eq:Gamma}
\end{equation}
where ${\bf x}$, is the three-dimensional WIMP position vector, $\epsilon({\bf x})$ is the spatial density of WIMPs in the Sun, and $\langle \sigma_{\rm ann} v_{\rm rel} \rangle\simeq3\times10^{-26}$~cm$^3$~s$^{-1}$ is the thermally averaged WIMP annihilation cross-section $\sigma_{\rm ann}$ times relative velocity $v_{\rm rel}$.~Consistently with Eq.~(\ref{eq:Gamma}), the probability of WIMP pair annihilation per unit time, $C_a$, can be written in terms of the annihilation rate $\Gamma_a$ as follows~$C_a=2\Gamma_a/N_\chi^2$, where $N_\chi(t)$ is the time dependent number of WIMPs trapped in the Sun at the time $t$.~The above expression for $C_a$ leads to the following relation between $C_a$ and $\langle \sigma_{\rm ann} v_{\rm rel} \rangle$
\begin{equation}
C_a = \langle \sigma_{\rm ann} v_{\rm rel} \rangle \frac{V_2}{V_1^2}\,,
\label{eq:CA}
\end{equation}
where 
\begin{equation}
\label{eq:vol}
V_1 = \int d^3{\bf x} \,\frac{\epsilon({\bf x})}{\epsilon_0} \,; \qquad \qquad V_2 = \int d^3{\bf x} \,\frac{\epsilon^2({\bf x})}{\epsilon_0^2}\,,
\end{equation}
and $\epsilon_0$ is the WIMP density at the Sun's centre.~Here, the time evolution of $N_\chi$ is governed by the differential equation
\begin{equation}
\dot{N}_\chi = C - C_a N_\chi^2 \,,
\label{eq:N}
\end{equation}
which admits the following solution
\begin{equation}
N_\chi(t) = \sqrt{\frac{C}{C_a}}\tanh\left(\sqrt{C C_a}t\right) \,,
\end{equation}
from which we find
\begin{equation}
\label{eq:annrate}
\Gamma_a=\frac{C}{2}\tanh^2\left(\sqrt{C C_a}t\right) \,.
\end{equation}
In the above equation, $\Gamma_a$ has to be evaluated at the present time, namely $t_\oplus=4.5\times 10^9$ years, which implies $\sqrt{C C_a}t_\oplus\ll 1$ and $\Gamma_a\simeq C/2$ for all DM interactions considered in this study~\cite{Widmark:2017yvd}.

The differential neutrino flux from WIMP annihilation in the Sun depends linearly on $\Gamma_a$, and is given by~\cite{Jungman:1995df}
\begin{equation}
    \frac{{\rm d} \Phi_\nu}{{\rm d} E_\nu} =
    \frac{\Gamma_a}{4\pi D^2}\sum_f B^f_\chi \frac{{\rm d} N^f_\nu}{{\rm d} E_\nu}\,,
\label{eq:nuflux}
\end{equation}
where $B^f_\chi$ is the branching ratio for WIMP pair annihilation into the final state $f$, ${\rm d} N^f_\nu / {\rm d} E_\nu$ is the neutrino energy spectrum at the detector from the decay of Standard Model particles in the final state $f$, $D$ is the detector's distance to the Sun's centre, and $E_\nu$ is the neutrino energy.

We compute the expected number of signal neutrinos using {\texttt{DARKSUSY}}~\cite{Bringmann:2018lay} to obtain the double differential flux of DM-induced neutrinos at the detector as a function of neutrino energy and polar angle, see Sec.~\ref{sec:nusignal}.~From now onwards, the term neutrinos refers to muon neutrinos.

\begin{table}[t]
    \centering
    \begin{tabular*}{\columnwidth}{@{\extracolsep{\fill}}llll@{}}
    \hline
        $\hat{\mathcal{O}}_1 = \mathds{1}_{\chi}\mathds{1}_N$  & $\hat{\mathcal{O}}_{10} = i{\bf{\hat{S}}}_N\cdot\frac{{\bf{\hat{q}}}}{m_N}\mathds{1}_\chi$   \\
        $\hat{\mathcal{O}}_3 = i{\bf{\hat{S}}}_N\cdot\left(\frac{{\bf{\hat{q}}}}{m_N}\times{\bf{\hat{v}}}^{\perp}\right)\mathds{1}_\chi$ & $\hat{\mathcal{O}}_{11} = i{\bf{\hat{S}}}_\chi\cdot\frac{{\bf{\hat{q}}}}{m_N}\mathds{1}_N$   \\
        $\hat{\mathcal{O}}_4 = {\bf{\hat{S}}}_{\chi}\cdot {\bf{\hat{S}}}_{N}$ & $\hat{\mathcal{O}}_{12} = {\bf{\hat{S}}}_{\chi}\cdot \left({\bf{\hat{S}}}_{N} \times{\bf{\hat{v}}}^{\perp} \right)$  \\
        $\hat{\mathcal{O}}_5 = i{\bf{\hat{S}}}_\chi\cdot\left(\frac{{\bf{\hat{q}}}}{m_N}\times{\bf{\hat{v}}}^{\perp}\right)\mathds{1}_N$ & $\hat{\mathcal{O}}_{13} =i \left({\bf{\hat{S}}}_{\chi}\cdot {\bf{\hat{v}}}^{\perp}\right)\left({\bf{\hat{S}}}_{N}\cdot \frac{{\bf{\hat{q}}}}{m_N}\right)$\\
        $\hat{\mathcal{O}}_6 = \left({\bf{\hat{S}}}_\chi\cdot\frac{{\bf{\hat{q}}}}{m_N}\right) \left({\bf{\hat{S}}}_N\cdot\frac{\hat{{\bf{q}}}}{m_N}\right)$ &  $\hat{\mathcal{O}}_{14} = i\left({\bf{\hat{S}}}_{\chi}\cdot \frac{{\bf{\hat{q}}}}{m_N}\right)\left({\bf{\hat{S}}}_{N}\cdot {\bf{\hat{v}}}^{\perp}\right)$  \\
        $\hat{\mathcal{O}}_7 = {\bf{\hat{S}}}_{N}\cdot {\bf{\hat{v}}}^{\perp}\mathds{1}_\chi$ &  $\hat{\mathcal{O}}_{15} = -\left({\bf{\hat{S}}}_{\chi}\cdot \frac{{\bf{\hat{q}}}}{m_N}\right)\left[ \left({\bf{\hat{S}}}_{N}\times {\bf{\hat{v}}}^{\perp} \right) \cdot \frac{{\bf{\hat{q}}}}{m_N}\right] $ \\
        $\hat{\mathcal{O}}_8 = {\bf{\hat{S}}}_{\chi}\cdot {\bf{\hat{v}}}^{\perp}\mathds{1}_N$ &  $\hat{\mathcal{O}}_{17}=i \frac{{\bf{\hat{q}}}}{m_N} \cdot \mathbf{\mathcal{S}} \cdot {\bf{\hat{v}}}^{\perp} \mathds{1}_N$ \\
        $\hat{\mathcal{O}}_9 = i{\bf{\hat{S}}}_\chi\cdot\left({\bf{\hat{S}}}_N\times\frac{{\bf{\hat{q}}}}{m_N}\right)$ & $\hat{\mathcal{O}}_{18}=i \frac{{\bf{\hat{q}}}}{m_N} \cdot \mathbf{\mathcal{S}}  \cdot {\bf{\hat{S}}}_{N}$  \\
    \hline
    \end{tabular*}
    \caption{Quantum mechanical operators defining the non-relativistic effective theory of DM-nucleon interactions~\cite{Fitzpatrick:2012ix}.~Here we do not consider the interaction operators $\hat{\mathcal{O}}_{2}$ and $\hat{\mathcal{O}}_{16}$, since the former is quadratic in $\mathbf{\hat{v}}^\perp$ (and we truncate the effective theory expansion at linear order in $\mathbf{\hat{v}}^\perp$ and second order in $\mathbf{\hat{q}}$) and the latter is a linear combination of $\hat{\mathcal{O}}_{12}$ and $\hat{\mathcal{O}}_{15}$.~The operators are expressed in terms of the basic invariants under Galilean transformations and Hermitian conjugation:~the identities in the nucleon and DM spin spaces, $\mathds{1}_{\chi}$ and $\mathds{1}_N$, the nucleon and DM spin operators, denoted by $\mathbf{\hat{S}}_N$ and $\mathbf{\hat{S}}_\chi$, respectively, the momentum transfer, $\mathbf{\hat{q}}$, and the transverse relative velocity operator $\mathbf{\hat{v}}^\perp$.~Here, $m_N$ is the nucleon mass, and all operators have the same mass dimension.~Standard spin-independent and spin-dependent interactions correspond to the operators $\hat{\mathcal{O}}_{1}$ and $\hat{\mathcal{O}}_{4}$, respectively.~The symbol $\mathbf{\mathcal{S}}$ denotes a symmetric combination of spin 1 polarisation vectors.~The operators $\hat{\mathcal{O}}_{17}$ and $\hat{\mathcal{O}}_{18}$ can only arise for spin 1 DM~\cite{Dent:2015zpa}.}
\label{tab:operators}
\end{table}

\subsection{Effective theory of dark matter-nucleon interactions}
\label{sec:eft}
Let us now focus on the cross section for DM-nucleus scattering, ${\rm d}\sigma_{\chi T}/{\rm d}E_r$.~We calculate  ${\rm d}\sigma_{\chi T}/{\rm d}E_r$ within the non-relativistic effective theory of DM-nucleon interactions~\cite{Fan:2010gt,Fitzpatrick:2012ix}.~The theory is built upon two assumptions:~1)~In the non-relativistic limit, the amplitude for dark matter scattering off nucleons in target nuclei can be expanded in powers of $|\mathbf{q}|/m_{N}\ll 1$, where  $m_N$ is the nucleon mass and $|\mathbf{q}|$ is the momentum transferred in the scattering.~2)~Each term in the above expansion is invariant under Galilean transformations and Hermitian conjugation.~As a result, it can be expressed in terms of $\mathbf{S_\chi}$, $\mathbf{S_N}$, $i\mathbf{q}$, and $\mathbf{v^\perp} \equiv \mathbf{v} + \mathbf{q}/2 \mu_N$, where $\mathbf{S_{\chi}}$ ($\mathbf{S_N}$) is the DM (nucleon) spin, and $\mu_N$ and $\mathbf{v}$ are the DM-nucleon reduced mass and relative velocity, respectively.~These, together with the identity, are the basic invariants under Galilean transformations and Hermitian conjugation.~At second order in the momentum transfer, and at first order in the transverse relative velocity $\mathbf{v^\perp}$, the effective theory of DM-nucleon interactions predicts 16 interaction operators, see Tab.~\ref{tab:operators}.~Here we only consider 14 of them, i.e.~all possible interactions that can arise for spin $\le1/2$ DM and are at most quadratic in the momentum transfer and linear in the transverse relative velocity, and neglect the operators denoted by $\hat{\mathcal{O}}_{17}$ and $\hat{\mathcal{O}}_{18}$~\cite{Dent:2015zpa}.~We refer to the original literature for further details concerning the construction of the effective theory of DM-nucleon interactions.~Within this theory, the differential cross section for DM-nucleus scattering, ${\rm d}\sigma_{T\chi}/{\rm d} E_r$, can be expressed as follows
\begin{align}
\frac{{\rm d}\sigma_{T\chi}\left(E_r,w^2\right)}{{\rm d}E_r} = \alpha\left(v^2\right) \hspace{-0.1 cm}\sum_{\tau,\tau',k}\left(\frac{q^2}{m_N^2}\right)^{\ell(k)} 
R^{\tau\tau'}_k\hspace{-0.15 cm}\left(v_T^{\perp 2}, {q^2 \over m_N^2} \right) W_k^{\tau\tau'}
\left(q^2\right)  \,,
\label{eq:dsigma} 
\end{align}
where $\alpha\left(v^2\right)=2m_T/[(2J_T+1)w^2]$, $m_T$ is the target nucleus mass, $J_T$ is the nuclear spin, $v_T^{\perp 2}=w^2-q^2/(4\mu_T^2)$, and $q=|\mathbf{q}|=\sqrt{2 m_T E_r}$.~Eight DM response functions, $R^{\tau\tau'}_k$, with $k=M,\Sigma',\Sigma'',\Phi'', \Phi'' M, \tilde{\Phi}', \Delta, \Delta \Sigma'$~\cite{Fitzpatrick:2012ix} appear in Eq.~(\ref{eq:dsigma}).~They are known analytically and depend on $v_T^{\perp 2}$, the expansion ``parameter'' $q^2/m_N^2$, and on the coupling constants for DM-nucleon interactions, $c_j^\tau$.~The index $j$ runs over all DM-nucleon interactions listed in Tab.~\ref{tab:operators}, while the indexes $\tau$ and $\tau'$ run from 0 to 1.~Here, $\tau=0$ refers to so-called ``isoscalar'' interactions and $\tau=$1 corresponds to ``isovector'' interactions.~In Eq.~(\ref{eq:dsigma}), $\ell(k)=0$ for $k=M,\Sigma',\Sigma''$, and $\ell(k)=1$ otherwise.~Explicitly, the DM response functions for spin $\le 1/2$ DM are given by~\cite{Fitzpatrick:2012ix} 
\begin{eqnarray}
 R_{M}^{\tau \tau^\prime}\left(v_T^{\perp 2}, {q^2 \over m_N^2}\right) &=& c_1^\tau c_1^{\tau^\prime } + {J_\chi (J_\chi+1) \over 3} \left[ {q^2 \over m_N^2} v_T^{\perp 2} c_5^\tau c_5^{\tau^\prime }+v_T^{\perp 2}c_8^\tau c_8^{\tau^\prime }
+ {q^2 \over m_N^2} c_{11}^\tau c_{11}^{\tau^\prime } \right] \nonumber \\
 R_{\Phi^{\prime \prime}}^{\tau \tau^\prime}\left(v_T^{\perp 2}, {q^2 \over m_N^2}\right) &=& {q^2 \over 4 m_N^2} c_3^\tau c_3^{\tau^\prime } + {J_\chi (J_\chi+1) \over 12} \left( c_{12}^\tau-{q^2 \over m_N^2} c_{15}^\tau\right) \left( c_{12}^{\tau^\prime }-{q^2 \over m_N^2}c_{15}^{\tau^\prime} \right)  \nonumber \\
 R_{\Phi^{\prime \prime} M}^{\tau \tau^\prime}\left(v_T^{\perp 2}, {q^2 \over m_N^2}\right) &=&  c_3^\tau c_1^{\tau^\prime } + {J_\chi (J_\chi+1) \over 3} \left( c_{12}^\tau -{q^2 \over m_N^2} c_{15}^\tau \right) c_{11}^{\tau^\prime } \nonumber \\
  R_{\tilde{\Phi}^\prime}^{\tau \tau^\prime}\left(v_T^{\perp 2}, {q^2 \over m_N^2}\right) &=&{J_\chi (J_\chi+1) \over 12} \left[ c_{12}^\tau c_{12}^{\tau^\prime }+{q^2 \over m_N^2}  c_{13}^\tau c_{13}^{\tau^\prime}  \right] \nonumber \\
   R_{\Sigma^{\prime \prime}}^{\tau \tau^\prime}\left(v_T^{\perp 2}, {q^2 \over m_N^2}\right)  &=&{q^2 \over 4 m_N^2} c_{10}^\tau  c_{10}^{\tau^\prime } +
  {J_\chi (J_\chi+1) \over 12} \left[ c_4^\tau c_4^{\tau^\prime} + \right.  \nonumber \\
 && \left. {q^2 \over m_N^2} ( c_4^\tau c_6^{\tau^\prime }+c_6^\tau c_4^{\tau^\prime })+
 {q^4 \over m_N^4} c_{6}^\tau c_{6}^{\tau^\prime } +v_T^{\perp 2} c_{12}^\tau c_{12}^{\tau^\prime }+{q^2 \over m_N^2} v_T^{\perp 2} c_{13}^\tau c_{13}^{\tau^\prime } \right] \nonumber \\
    R_{\Sigma^\prime}^{\tau \tau^\prime}\left(v_T^{\perp 2}, {q^2 \over m_N^2}\right)  &=&{1 \over 8} \left[ {q^2 \over  m_N^2}  v_T^{\perp 2} c_{3}^\tau  c_{3}^{\tau^\prime } + v_T^{\perp 2}  c_{7}^\tau  c_{7}^{\tau^\prime }  \right]
       + {J_\chi (J_\chi+1) \over 12} \left[ c_4^\tau c_4^{\tau^\prime} +  \right.\nonumber \\
       &&\left. {q^2 \over m_N^2} c_9^\tau c_9^{\tau^\prime }+{v_T^{\perp 2} \over 2} \left(c_{12}^\tau-{q^2 \over m_N^2}c_{15}^\tau \right) \left( c_{12}^{\tau^\prime }-{q^2 \over m_N^2}c_{15}^{\tau \prime} \right) +{q^2 \over 2 m_N^2} v_T^{\perp 2}  c_{14}^\tau c_{14}^{\tau^\prime } \right] \nonumber \\
     R_{\Delta}^{\tau \tau^\prime}\left(v_T^{\perp 2}, {q^2 \over m_N^2}\right)&=&  {J_\chi (J_\chi+1) \over 3} \left[ {q^2 \over m_N^2} c_{5}^\tau c_{5}^{\tau^\prime }+ c_{8}^\tau c_{8}^{\tau^\prime } \right] \nonumber \\
 R_{\Delta \Sigma^\prime}^{\tau \tau^\prime}\left(v_T^{\perp 2}, {q^2 \over m_N^2}\right)&=& {J_\chi (J_\chi+1) \over 3} \left[c_{5}^\tau c_{4}^{\tau^\prime }-c_8^\tau c_9^{\tau^\prime} \right]\,,
 \label{eq:R}
\end{eqnarray}
where we denote by $J_\chi$ the DM particle spin.~In Eq.~(\ref{eq:dsigma}), the eight nuclear response functions $W_k^{\tau\tau'}$, $k=M,\Sigma',\Sigma'',\Phi'', \Phi'' M, \tilde{\Phi}', \Delta, \Delta \Sigma'$ are quadratic in reduced matrix elements of nuclear charges and currents, and must be computed numerically.~For the 16 most abundant elements in the Sun, we use nuclear response functions obtained through nuclear shell model calculations in~\cite{Catena:2015uha}.~In our numerical computations, we will set the DM particle spin to $J_\chi=1/2$.

\section{Analysis}
\label{sec:analysis}
The main purpose of this article is to assess the sensitivity of a PINGU-like detector to the isoscalar and isovector component of the DM-nucleon interactions in Tab.~\ref{tab:operators}.~In this section we describe our analysis strategy, concentrating on statistical methods (Sec.~\ref{sec:stat}), neutrino signal (Sec.~\ref{sec:nusignal}), neutrino background (Sec.~\ref{sec:nuback}), and assumed effective area and angular resolution (Sec.~\ref{sec:exp}), separately.

\subsection{Statistical methods}
\label{sec:stat}
Assessing PINGU's sensitivity, we test the background-only hypothesis against the background~\hspace{-0.17cm}-plus-signal hypothesis using the standard likelihood ratio method.~For a counting experiment with a sufficiently large number of background events, this yields the following asymptotic result for the significance $S$ with which a point in the DM mass -- coupling constant plane can be observed~\cite{Cowan:2010js}~\footnote{Here, the approximation in the last equality in Eq.~(\ref{eq:sign1}) induces a relative error on $S$ of at most 20\%.}
\begin{equation} \label{eq:sign1}
S=\sqrt{2\big((N_s+N_{b})\log\big( 1+\dfrac{N_s}{N_{b}}\big)-N_s \big)}\approx\dfrac{N_{s}}{\sqrt{N_{b}}},
\end{equation}
where $N_s$ is the number of signal muon neutrinos and $N_{b}$ is the number of background muon neutrinos times a veto factor to account for atmospheric muon contamination (see Sec.~\ref{sec:nuback} for further details on how we compute $N
_b$).~In the last step in Eq.~(\ref{eq:sign1}), we assumed $N_s \ll N_b$, consistently with previous background estimations~\cite{Honda}.~The significance, $S$, is related to the $p$-value via
\begin{equation}
S=\Phi^{-1}(1-p),
\label{eq:sign2}
\end{equation}
where $\Phi^{-1}$ is the inverse cumulative distribution of a Gaussian function with mean 0 and variance 1, i.e.~the so-called quantile.~In Sec.~\ref{sec:results}, we will use Eq.~(\ref{eq:sign2}) to compare current exclusion limits on DM-nucleon coupling constants to the projected sensitivity that we find for PINGU.~Specifically, we will compare 90\% confidence level (C.L.) exclusion limits corresponding to $S=1.28$, to $5\sigma$ sensitivity projections corresponding to $S=5$.~Strictly speaking, this comparison is only accurate when the significance for exclusion coincides with the significance for discovery~\cite{Cowan:2010js}.

\subsection{Neutrino signal}
\label{sec:nusignal}
\begin{figure}[t]
    \centering
    \includegraphics[width=0.6\linewidth]{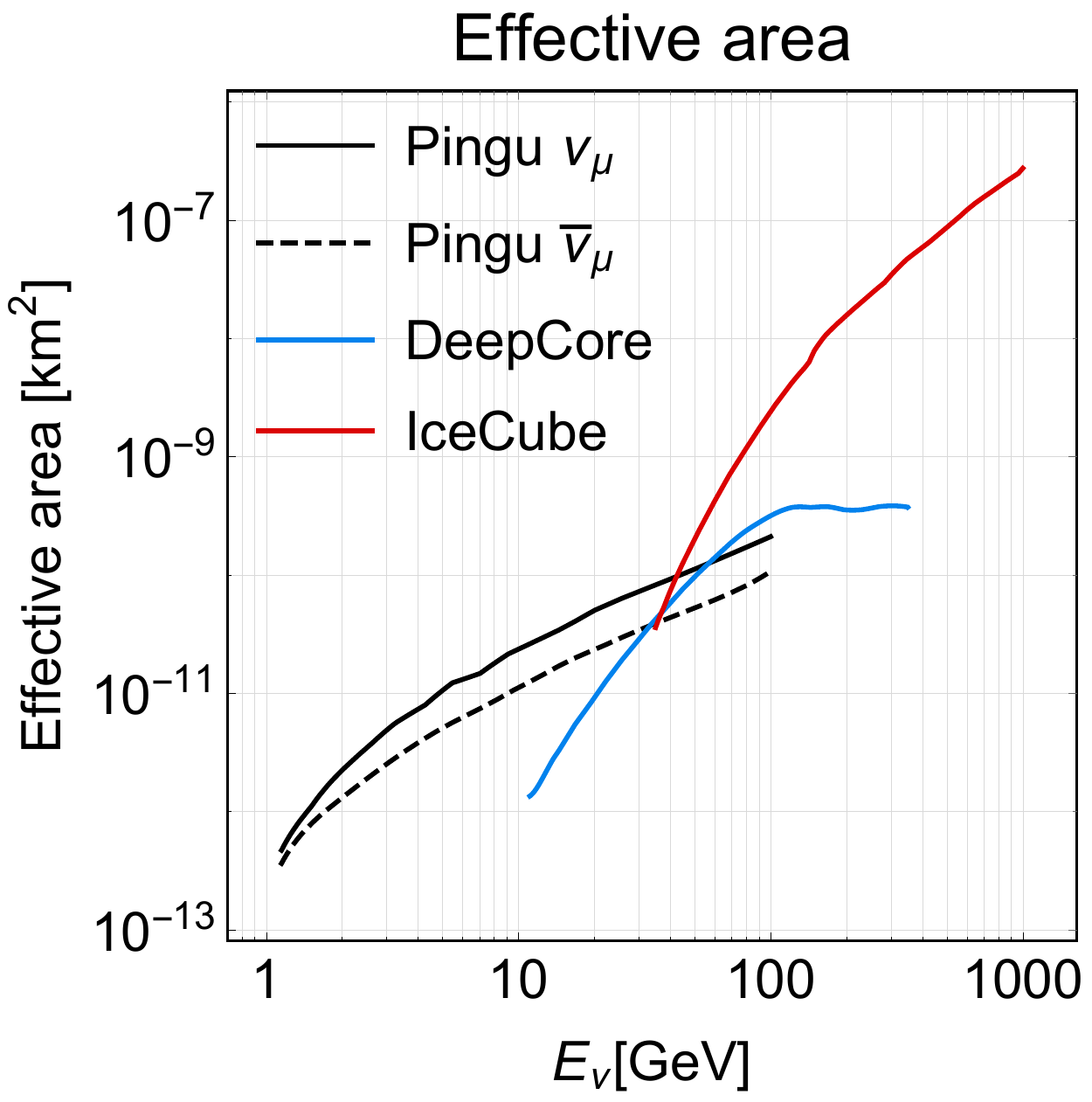}
    \caption{\emph{ Muon neutrino effective areas used in this analysis. For IceCube (red line) and DeepCore (blue line) we plot the total effective areas from~\cite{IC3}.~For PINGU (black lines), we show $A^{\rm eff}_{\nu_\mu}$ and $A^{\rm eff}_{\bar{\nu}_\mu}$ separately (see \cite{LoI} and Eq.~(\ref{eq:effpingu})).}
    }
    \label{EffArea}
\end{figure}

In order to calculate the number of signal neutrino events at PINGU (and, for comparison, at IceCube), we proceed as follows.~First, we calculate the rate of DM capture in the Sun using a modified version of the \texttt{MATHEMATICA} package \texttt{DMFormFactor}~\cite{Anand:2013yka} which implements Eq.~(\ref{eq:C}) and the nuclear response functions found in~\cite{Catena:2015uha} for the 16 most abundant elements in the Sun.~We then use the capture rates found in this way as an input for the {\texttt{DARKSUSY}} code~\cite{Bringmann:2018lay}, from which we extract the double differential muon neutrino flux at the detector, ${\rm d}\Phi_{\nu_\mu}/({\rm d}E_{\nu_\mu}{\rm d}\Omega)$, and the corresponding antineutrino flux, ${\rm d}\Phi_{\bar{\nu}_\mu}/({\rm d}E_{\bar{\nu}_\mu}{\rm d}\Omega)$, where energies, $E_\nu$ and $E_{\bar{\nu}_\mu}$, and solid angle ${\rm d}\Omega={\rm d}\hspace{-0.07cm}\cos\hspace{-0.05cm}\theta{\rm d}\phi$ are defined below Eq.~(\ref{eq:signal}).~Finally, we convolve the double differential neutrino and antineutrino fluxes with the corresponding detector effective areas, $A^{\rm eff}_{\nu_\mu}$ and $A^{\rm eff}_{\bar{\nu}_\mu}$, respectively, obtaining~\footnote{For IceCube/DeepCore, $A^{\rm eff}_{\nu_\mu} = A^{\rm eff}_{\bar{\nu}_\mu}=A^{\rm eff}$, and $A^{\rm eff}$ is given in Fig.~\ref{EffArea}.}
\begin{equation} \label{eq:signal}
N_{s}=\Delta t\sum_{\nu=\nu_\mu, \bar{\nu}_{\mu}}\int_{E_{\rm th}}^{m_\chi}{\rm d}E_{\nu} \int_0^{2\pi}{\rm d}\varphi \int_0^{\theta_{\rm res}(E_{\nu})}{\rm d}\theta\sin\theta A_{\nu}^{\mathrm{eff}}(E_{\nu})\dfrac{{\rm d}\Phi_{\nu}}{{\rm d}E_{\nu}{\rm d}\Omega}\,,
\end{equation}
where $\theta$ and $\varphi$ are polar and azimuthal angle in a reference frame with centre at the detector location and $z$-axis pointing in the Sun's centre direction, $E_{\nu_{\mu}}$ ($E_{\bar{\nu}_{\mu}}$) is the incoming neutrino (antineutrino) energy, $E_{\rm th}$ is the detector's energy threshold (which we extract from Fig.~\ref{EffArea}), $\theta_{\mathrm{res}}$ is the energy dependent median detector angular resolution~\cite{IC3}, and, finally, $\Delta t$ is the time of data taking (e.g.~532 days for IceCube/DeepCore).~Further details on $\theta_{\rm res}$ and $A^{\rm eff}_\nu$ are provided in Sec.~\ref{sec:exp}.

\subsection{Neutrino background}
\label{sec:nuback}
The neutrino background consists of atmospheric neutrinos from cosmic ray interactions in the atmosphere, and solar neutrinos from fusion processes inside the Sun.~Compared to neutrinos from WIMP annihilation, solar neutrinos are less energetic.~Furthermore, their expected rate at the detector is of about $1$ event per year~\cite{IC3}.~For these reasons, solar neutrinos will not be considered here.~Atmospheric neutrinos divide into a {\it prompt flux} (from atmospheric charm production by cosmic rays~\cite{Enberg:2008te}) and a {\it conventional flux} (from the decay of charged pions and kaons).~The prompt flux is roughly a thousand times smaller than the conventional flux at energies in the WIMP mass range and it will therefore be neglected~\cite{ConvFlux}.~For the South Pole, the conventional atmospheric neutrino flux has been measured and tabulated by Honda {\it et al.}~\cite{Honda}.~In this analysis, we use the results published in~\cite{Honda} after averaging the collected data over polar angle and time.

In addition to muon neutrinos, atmospheric muons constitute a second type of experimental background at neutrino telescopes, since muon neutrinos are detected by their conversion into muons via charged current interactions in ice.~Therefore, muon neutrinos from WIMP annihilation and atmospheric muons from cosmic rays interactions can be misidentified at neutrino telescopes.~At IceCube, misidentification of atmospheric muons is reduced by rejecting (or vetoing) events that start at the outer strings or events that show activity in the IceTop detector at the surface of the ice.~These events are more likely to be caused by atmospheric muons rather than muon neutrinos~\cite{LoI}.~Furthermore, only data collected when the Sun is below the horizon are considered, as in this case the Earth is used as a filter for atmospheric muons from the direction of the Sun~\cite{IC3,Icecube1}.~This occurs in Winter, when the local zenith angle is $\geq 90^\circ$.~We assume here that veto techniques in PINGU will be good enough to allow for data taking throughout the whole year~\cite{IC3}.~In practice, atmospheric muon contamination is taken into account by multiplying $N_b$ by a correction, or veto factor, $\eta > 1$.~We assume that $\eta$ does not depend on energy, and use the same value of $\eta$ for IceCube/DeepCore and PINGU.~For DeepCore, we estimate $\eta$ as follows.~We compute the expected number of muon neutrino events in the DeepCore detector during the time in which the Sun is below the horizon, integrating Honda's flux in a cone of opening angle $10^{\degree}$ around the Sun~\cite{IC3,Icecube1}.~We find 243 expected muon neutrino events from this region during the lifetime of the analysis presented in~\cite{Icecube1} whereas -- including muon contamination -- the IceCube Collaboration finds 428 neutrino events from the same region.~Accordingly, for the veto factor we use the estimate $\eta=428/243\simeq 1.76$.~This estimate is very close to the one reported by the IceCube collaboration, i.e.~$\eta\simeq 1.52$.~As already anticipated, we assume, conservatively, that the veto factor $\eta\simeq 1.76$ also applies to the PINGU experiment.

Taking into account the misidentification of atmospheric muons, we compute the expected number of atmospheric muon neutrino events, $N_{b}$, in a time interval $\Delta t$, using the following expression 
\begin{equation} \label{eq:bg}
N_{b}= \eta \Delta t\cdot 2\pi\sum_{\nu=\nu_\mu, \bar{\nu}_{\mu}}\int_{E_\mathrm{th}}^{m_\chi}{\rm d}E_{\nu} \left[1-\cos\theta_{\mathrm{res}}\left(E_{\nu}\right)\right]A_{\nu}^{\mathrm{eff}}(E_{\nu})\left\langle \dfrac{{\rm d}\Phi_\nu^{\mathrm{atm}}}{{\rm d}E_{\nu}{\rm d}\Omega}\right\rangle \,,
\end{equation}
where the notation is the same as in Eq.~(\ref{eq:signal}), while $\langle {\rm d}\Phi_\nu^{\mathrm{atm}}/{\rm d}E_{\nu}{\rm d}\Omega\rangle $ is the double differential muon neutrino flux from~\cite{HondaReport} averaged over time and solid angle.~We assume $\Delta t=532$ days of data taking with the Sun below the horizon for IceCube/DeepCore and $\Delta t=3$ years of continuous data taking for PINGU.~Atmospheric muon contamination is taken into account through $\eta$.

\begin{figure}[t]
\begin{center}
\begin{minipage}[t]{0.495\linewidth}
\centering
\includegraphics[width=\textwidth]{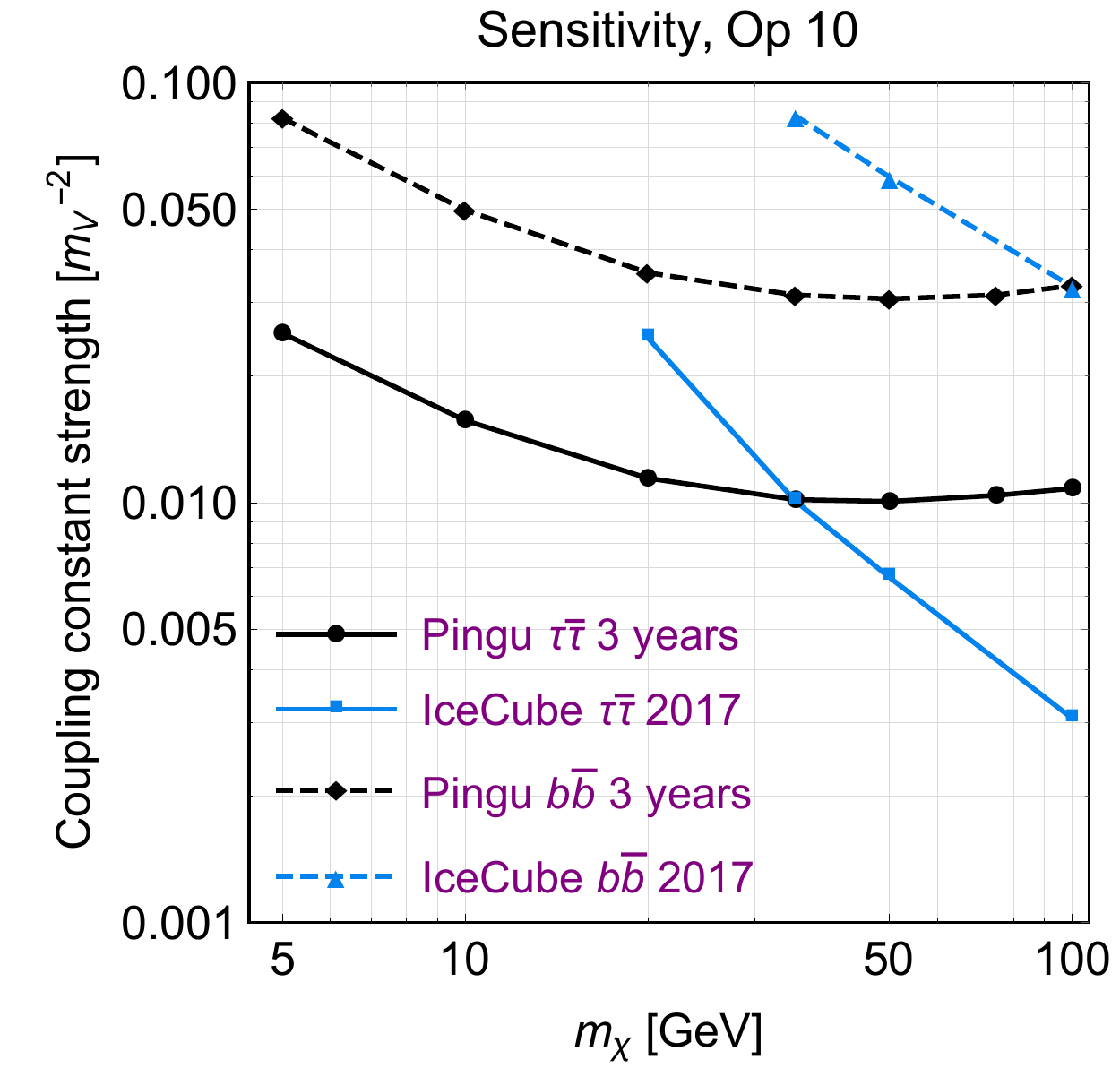}
\end{minipage}
\begin{minipage}[t]{0.495\linewidth}
\centering
\includegraphics[width=\textwidth]{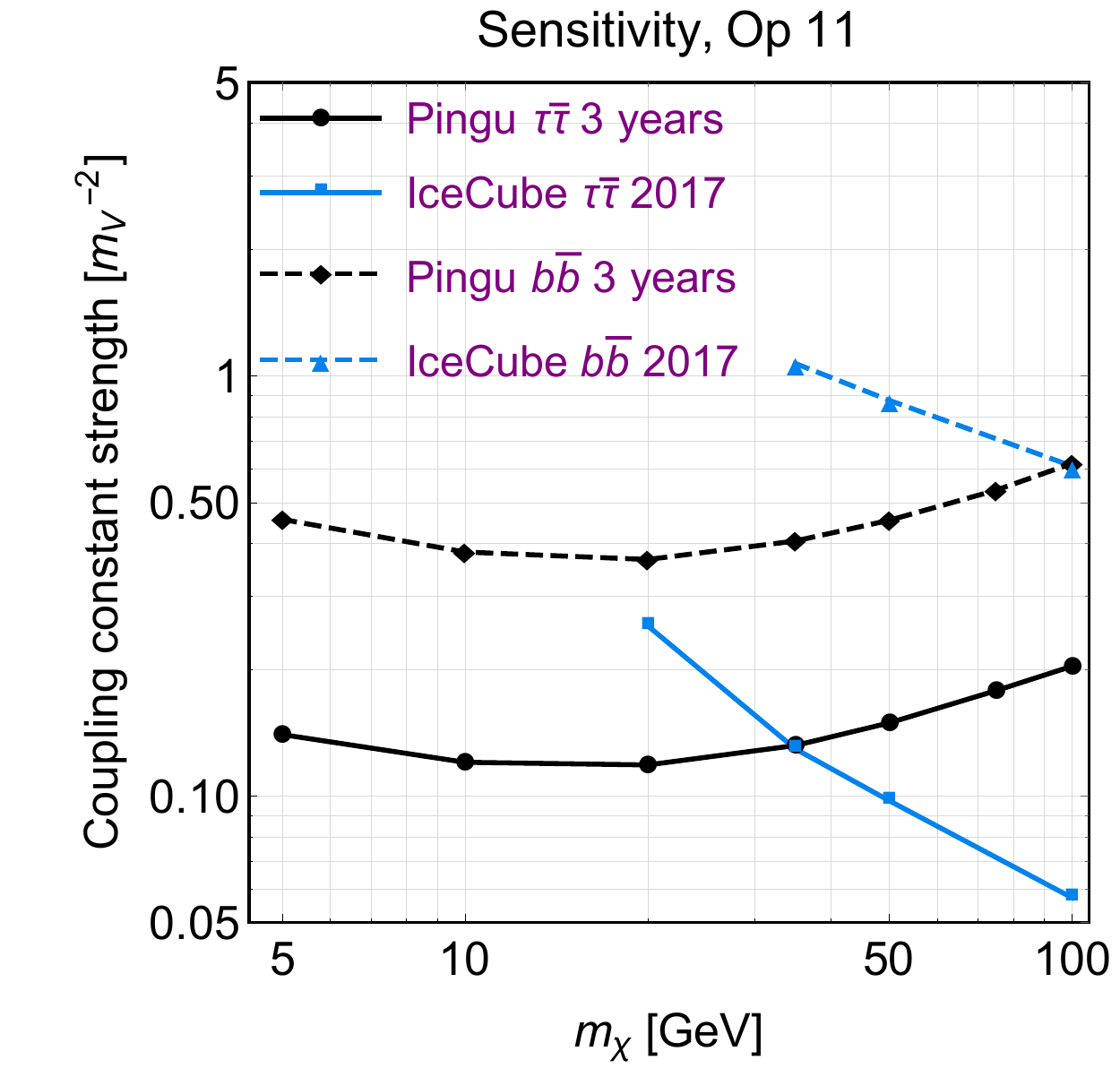}
\end{minipage}
\begin{minipage}[t]{0.495\linewidth}
\centering
\includegraphics[width=\textwidth]{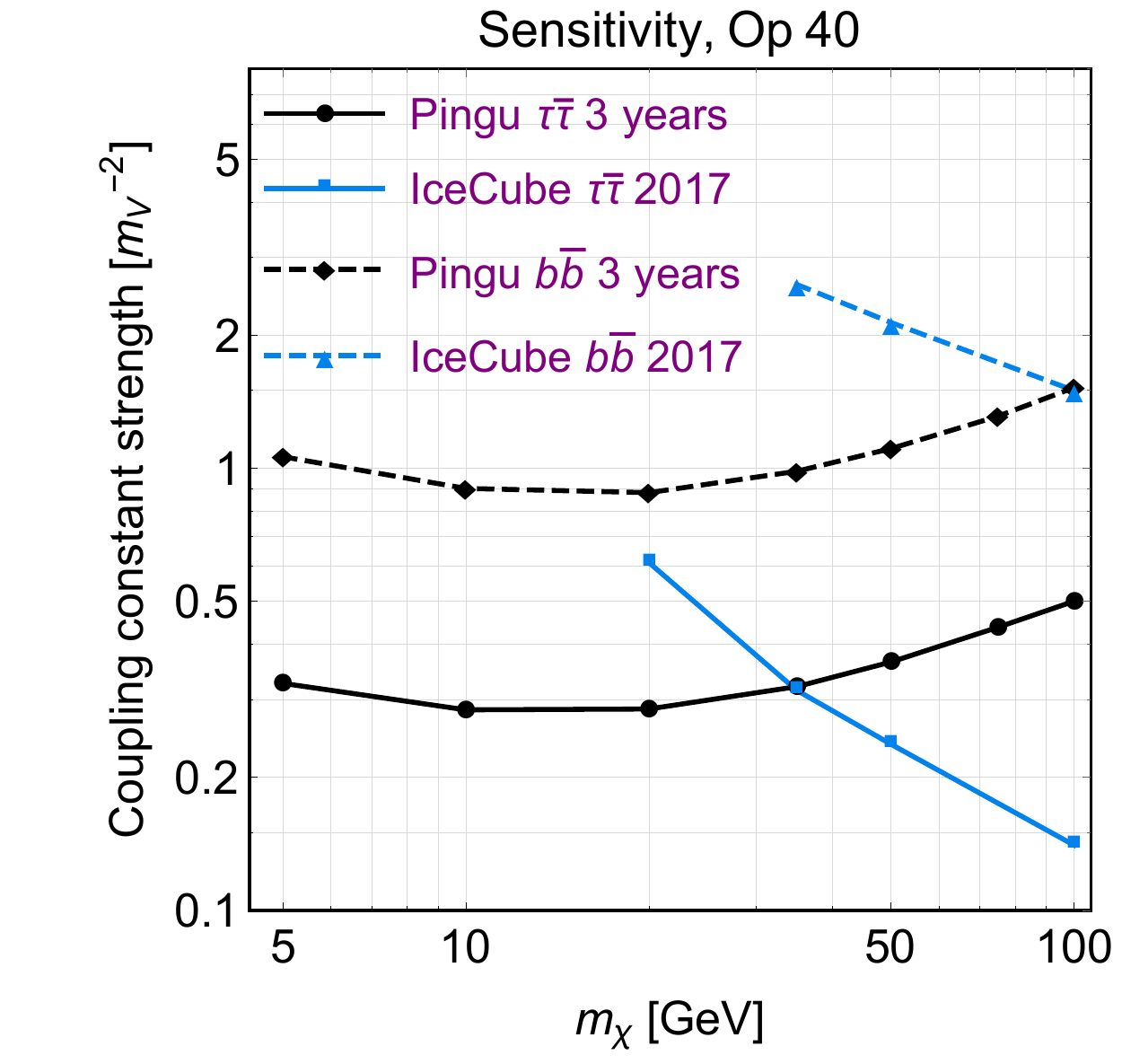}
\end{minipage}
\begin{minipage}[t]{0.495\linewidth}
\centering
\includegraphics[width=\textwidth]{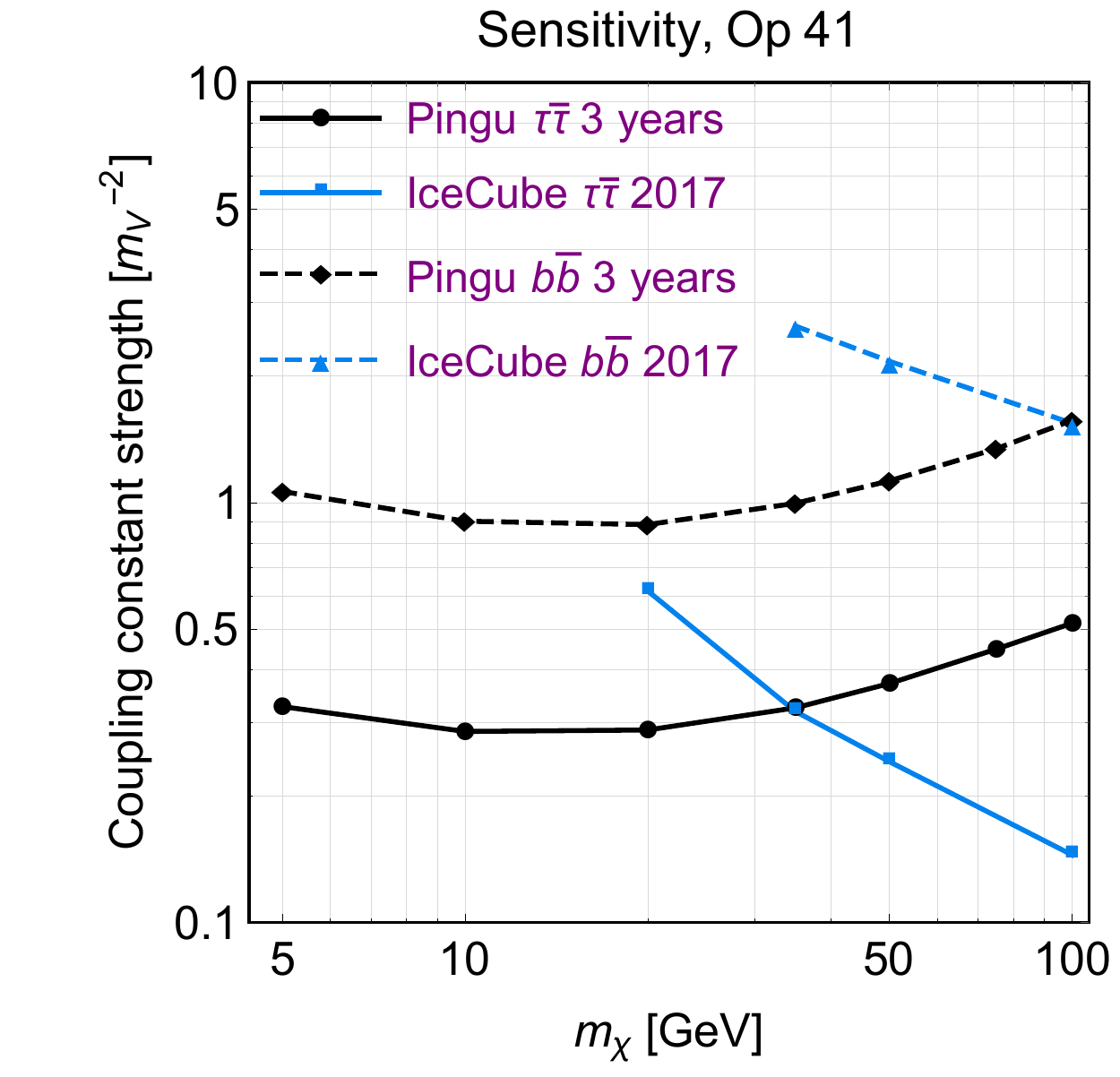}
\end{minipage}
\end{center}
\caption{PINGU's $5\sigma$ sensitivity contours (corresponding to a 5$\sigma$ significance) and 90\% C.L. exclusion limits from the null result of DM searches at IceCube/DeepCore for the interaction operators $\hat{\mathcal{O}}_1$ (canonical spin-independent interaction) and $\hat{\mathcal{O}}_4$ (canonical spin-dependent interaction).~Sensitivity contours  (black) and exclusion limits (light blue) are presented in the DM mass -- coupling constant plane.~Left panels refer to isoscalar interactions (Op10 and Op40) while right panels correspond to isovector interactions (Op11 and Op41).~Solid (dashed) lines assume $\tau\bar{\tau}$ ($b\bar{b}$) as the leading DM annihilation channel.~On the $y$-axis, coupling constants are expressed in units of the electroweak scale, i.e.~$m_V=246.2$~GeV.}  
\label{fig:op1op4}
\end{figure}

\begin{figure}[t]
\begin{center}
\begin{minipage}[t]{0.32\linewidth}
\centering
\includegraphics[width=\textwidth]{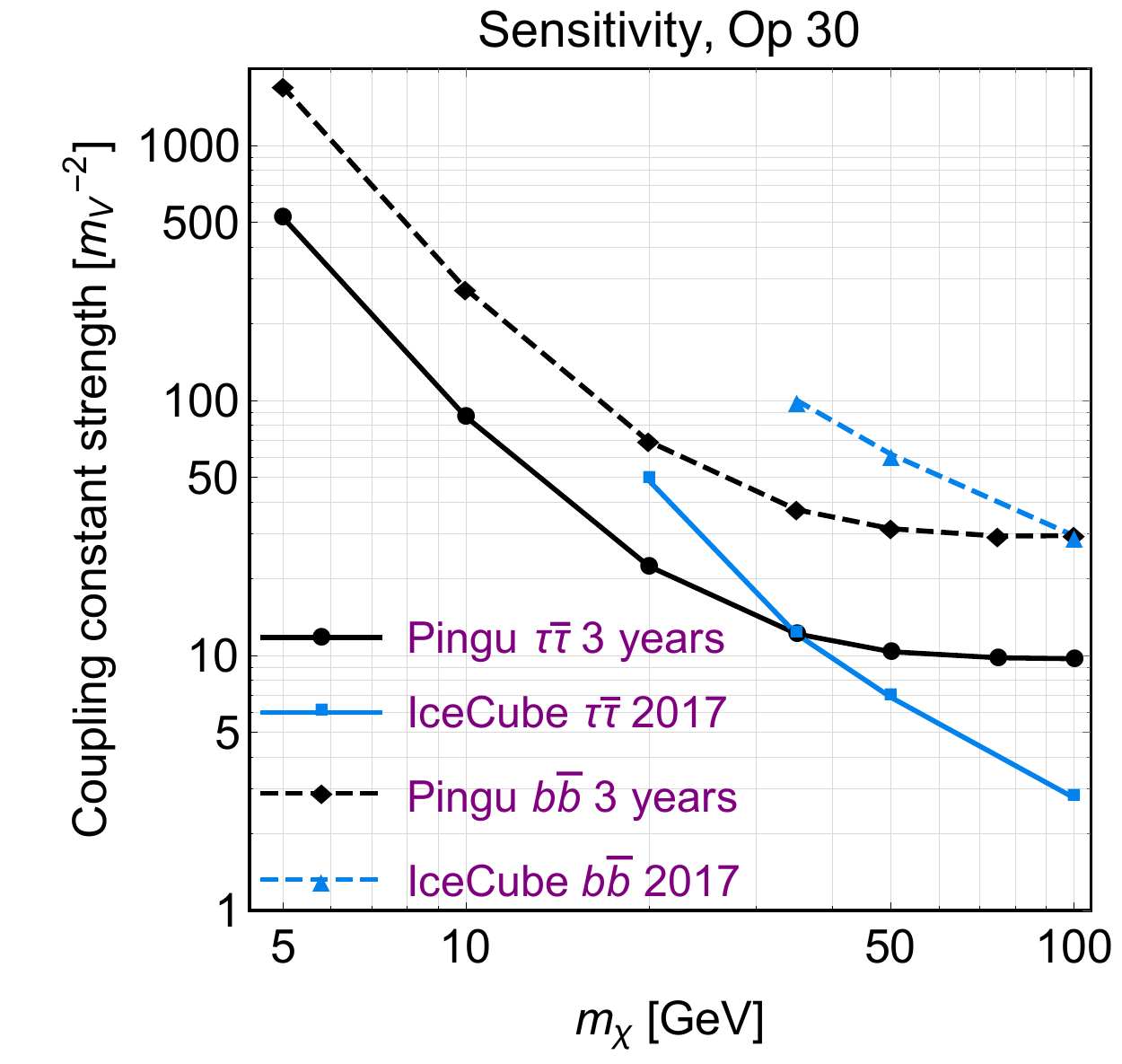}
\end{minipage}
\begin{minipage}[t]{0.32\linewidth}
\centering
\includegraphics[width=\textwidth]{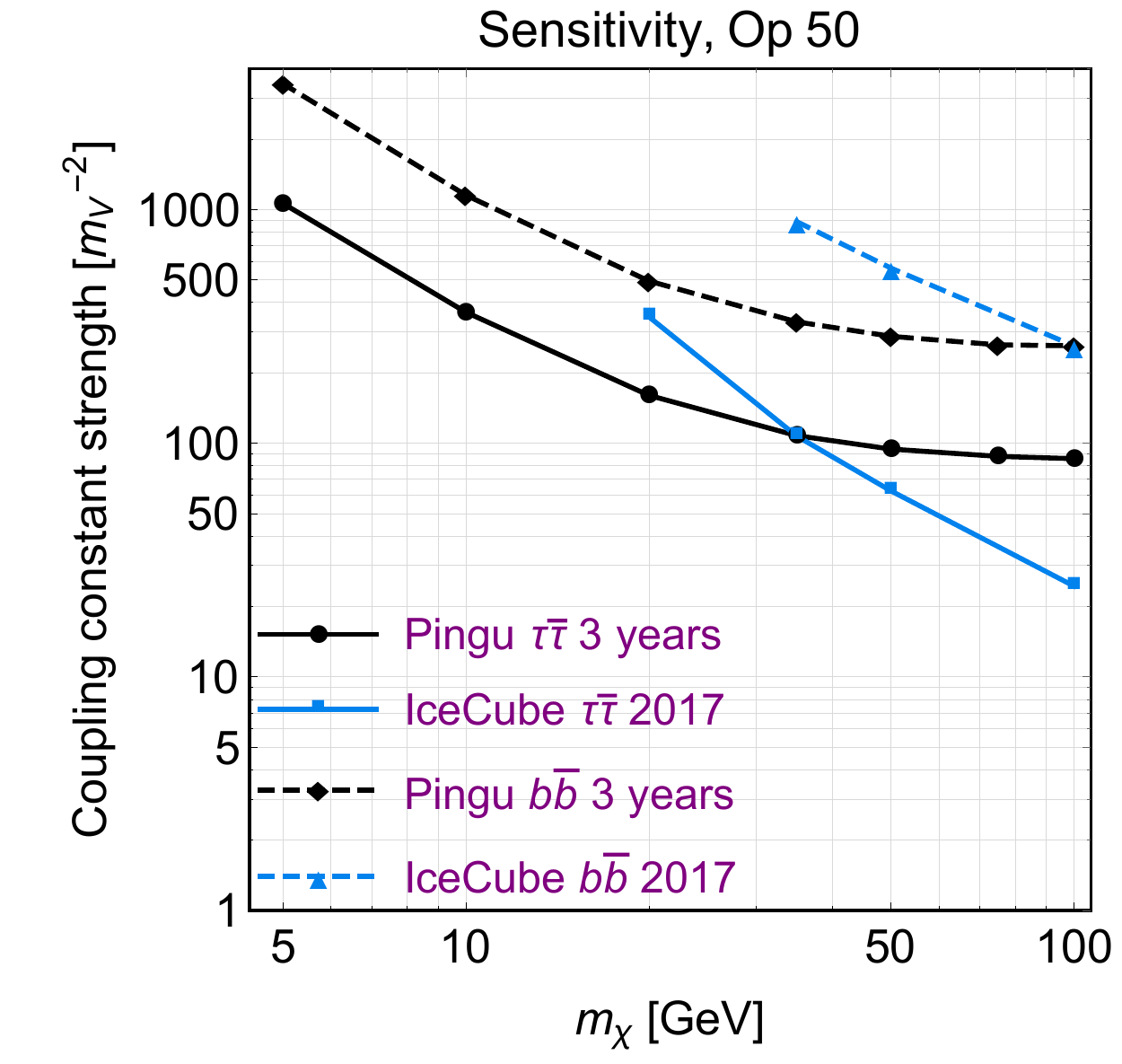}
\end{minipage}
\begin{minipage}[t]{0.32\linewidth}
\centering
\includegraphics[width=\textwidth]{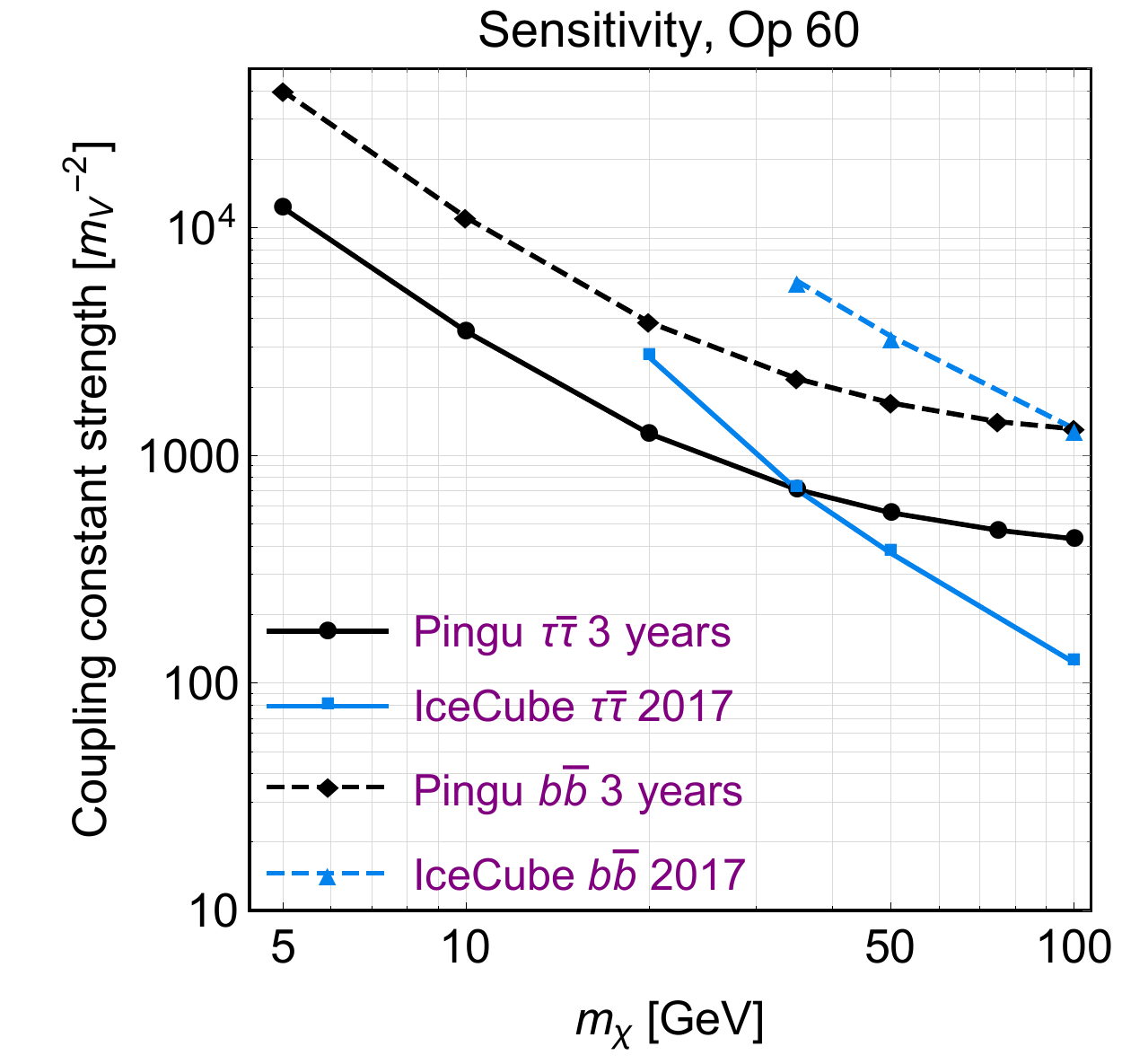}
\end{minipage}
\begin{minipage}[t]{0.32\linewidth}
\centering
\includegraphics[width=\textwidth]{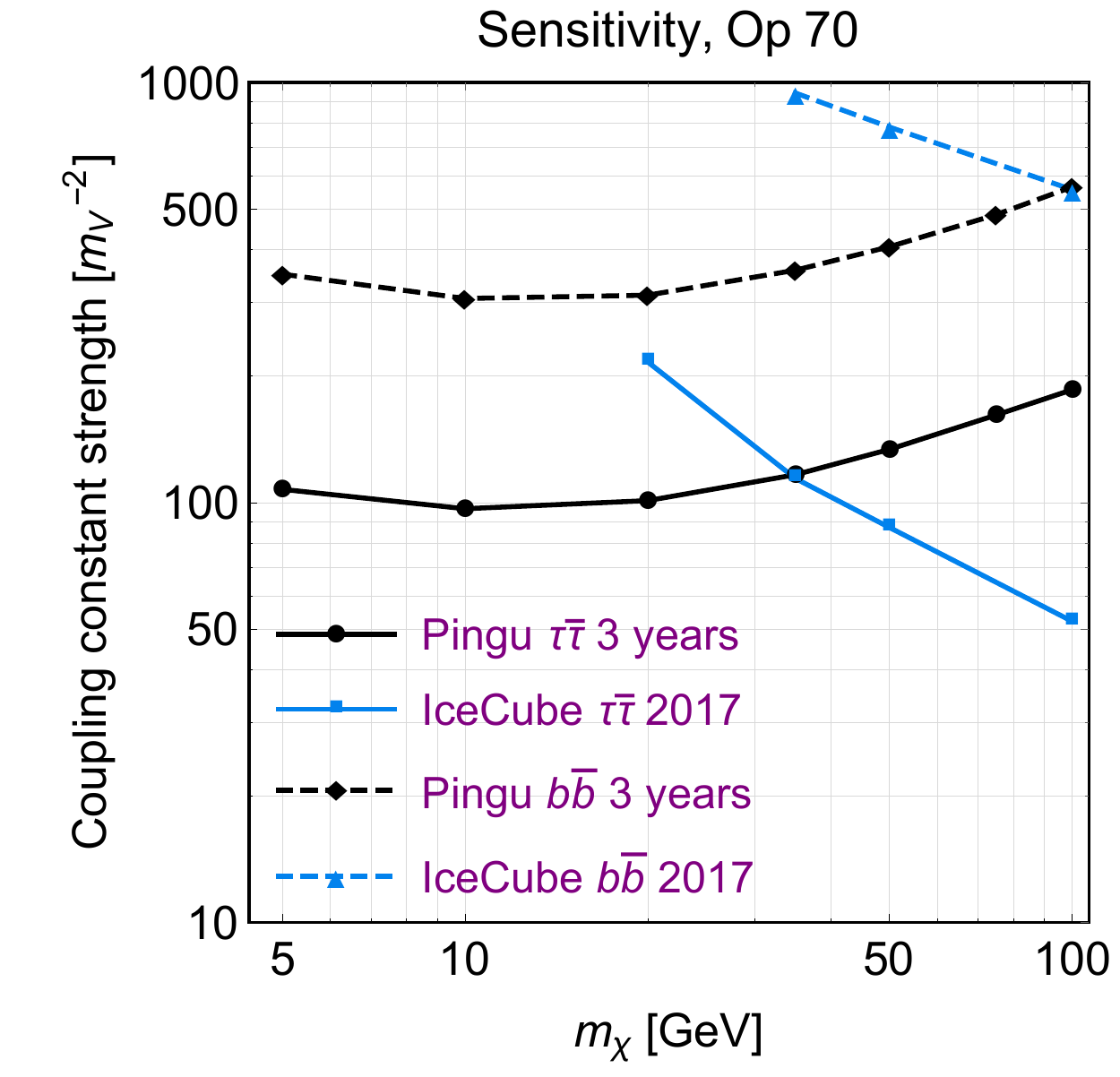}
\end{minipage}
\begin{minipage}[t]{0.32\linewidth}
\centering
\includegraphics[width=\textwidth]{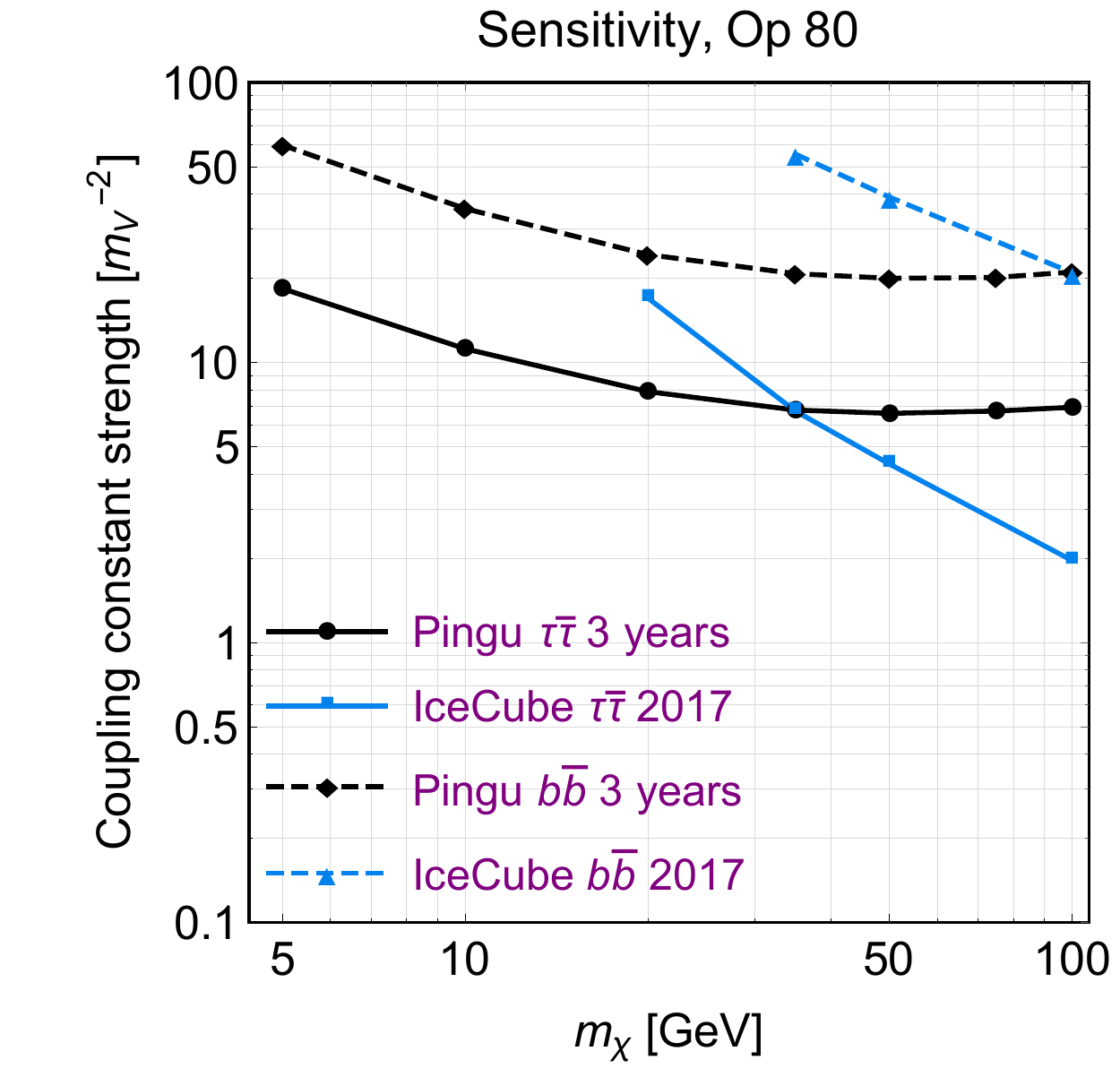}
\end{minipage}
\begin{minipage}[t]{0.32\linewidth}
\centering
\includegraphics[width=\textwidth]{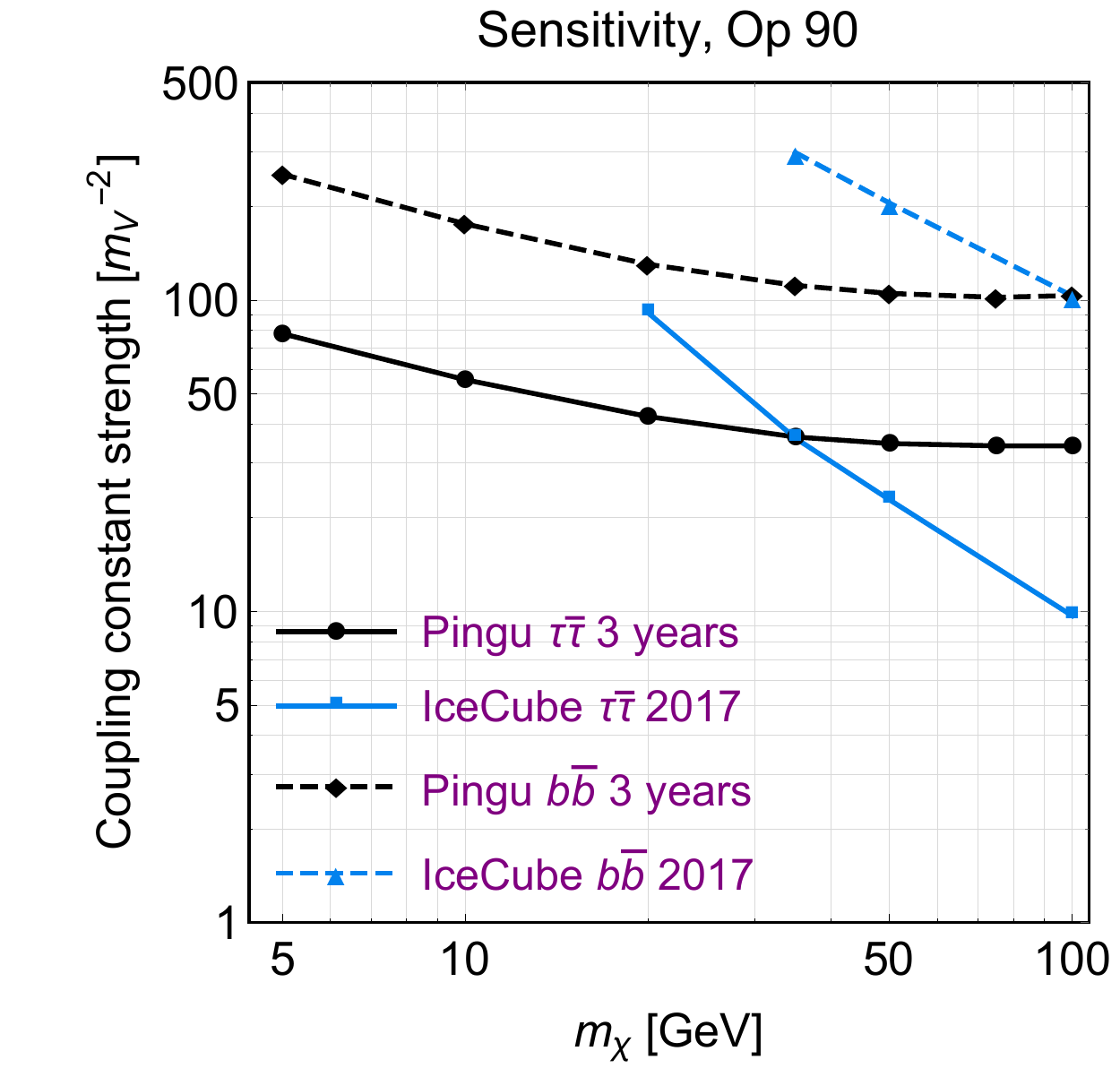}
\end{minipage}
\end{center}
\caption{Same as Fig.~\ref{fig:op1op4}, but now for the the isoscalar component of the operators $\hat{\mathcal{O}}_3=i{\bf{\hat{S}}}_N\cdot({\bf{\hat{q}}}/m_N)\times{\bf{\hat{v}}}^{\perp}\mathds{1}_\chi$, $\hat{\mathcal{O}}_5=i{\bf{\hat{S}}}_\chi\cdot[({{\bf{\hat{q}}}/m_N})\times{\bf{\hat{v}}}^{\perp}]\mathds{1}_N$, $\hat{\mathcal{O}}_6={\bf{\hat{S}}}_\chi\cdot({\bf{\hat{q}}}/m_N) {\bf{\hat{S}}}_N\cdot(\hat{{\bf{q}}}/m_N)$, $\hat{\mathcal{O}}_7={\bf{\hat{S}}}_{N}\cdot {\bf{\hat{v}}}^{\perp}\mathds{1}_\chi$, $\hat{\mathcal{O}}_8={\bf{\hat{S}}}_{\chi}\cdot {\bf{\hat{v}}}^{\perp}\mathds{1}_N$, and $\hat{\mathcal{O}}_9=i{\bf{\hat{S}}}_\chi\cdot[{\bf{\hat{S}}}_N\times({\bf{\hat{q}}}/m_N)]$.}  
\label{fig:isos1}
\end{figure}

\begin{figure}[t]
\begin{center}
\begin{minipage}[t]{0.32\linewidth}
\centering
\includegraphics[width=\textwidth]{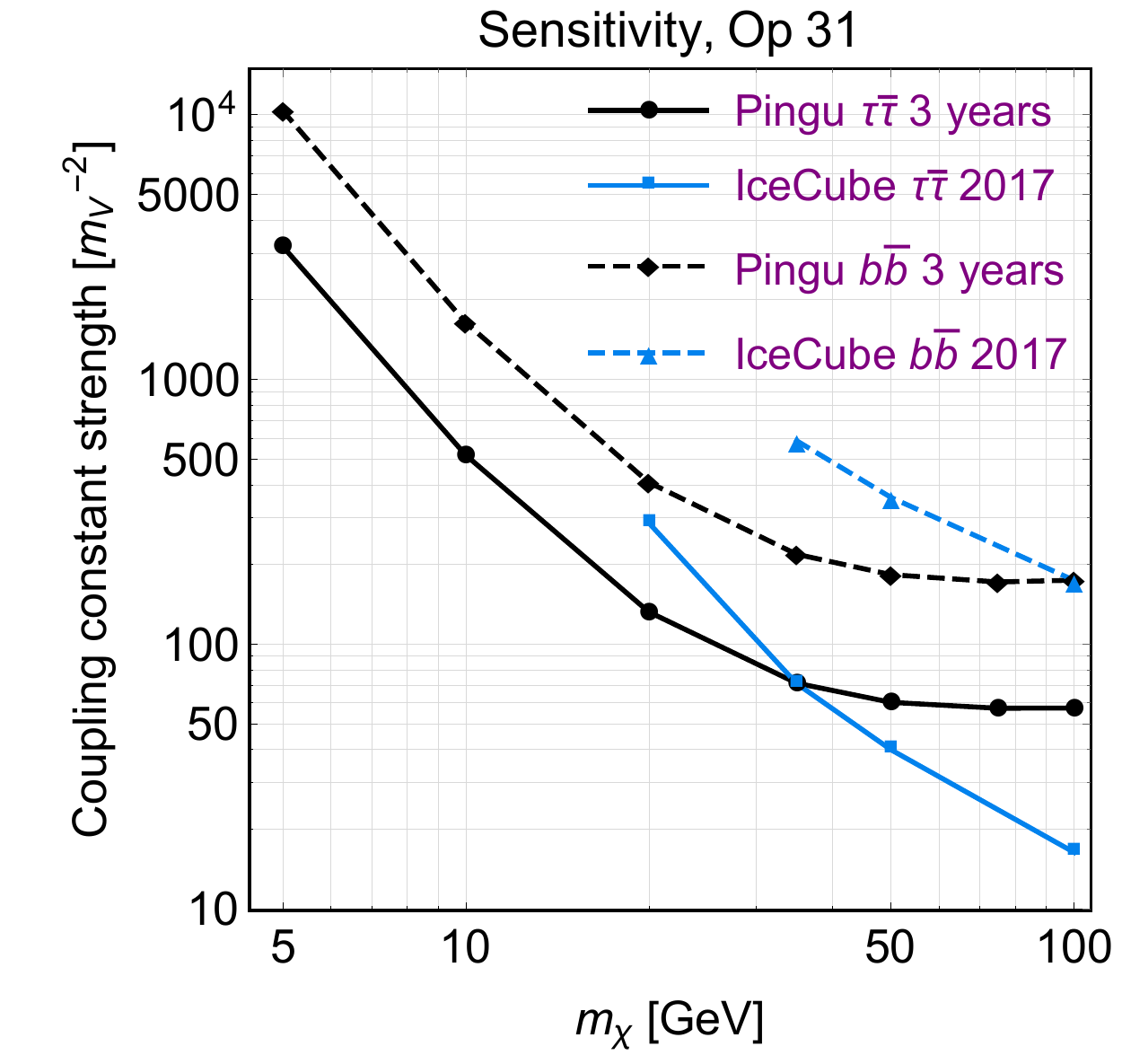}
\end{minipage}
\begin{minipage}[t]{0.32\linewidth}
\centering
\includegraphics[width=\textwidth]{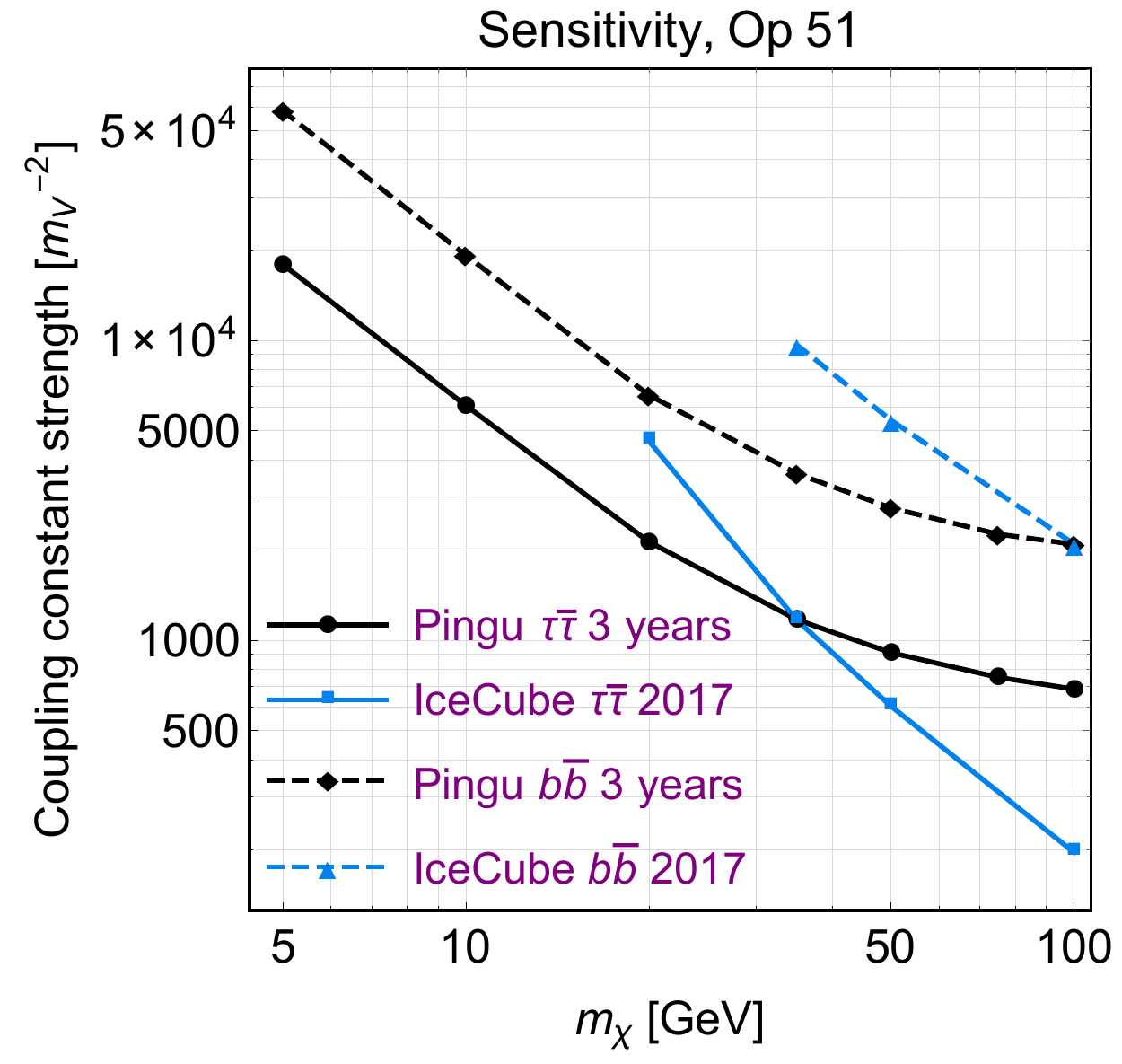}
\end{minipage}
\begin{minipage}[t]{0.32\linewidth}
\centering
\includegraphics[width=\textwidth]{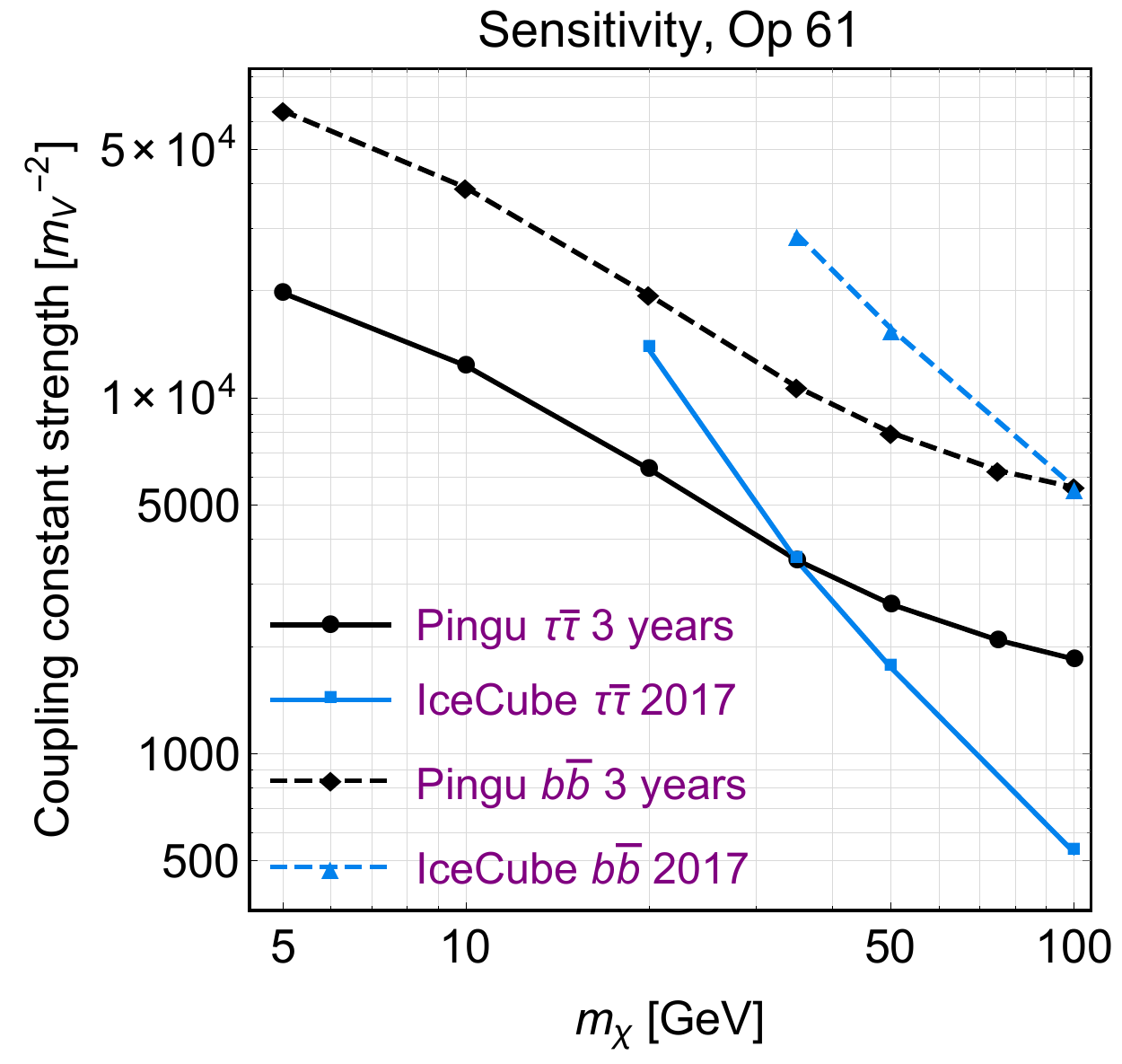}
\end{minipage}
\begin{minipage}[t]{0.32\linewidth}
\centering
\includegraphics[width=\textwidth]{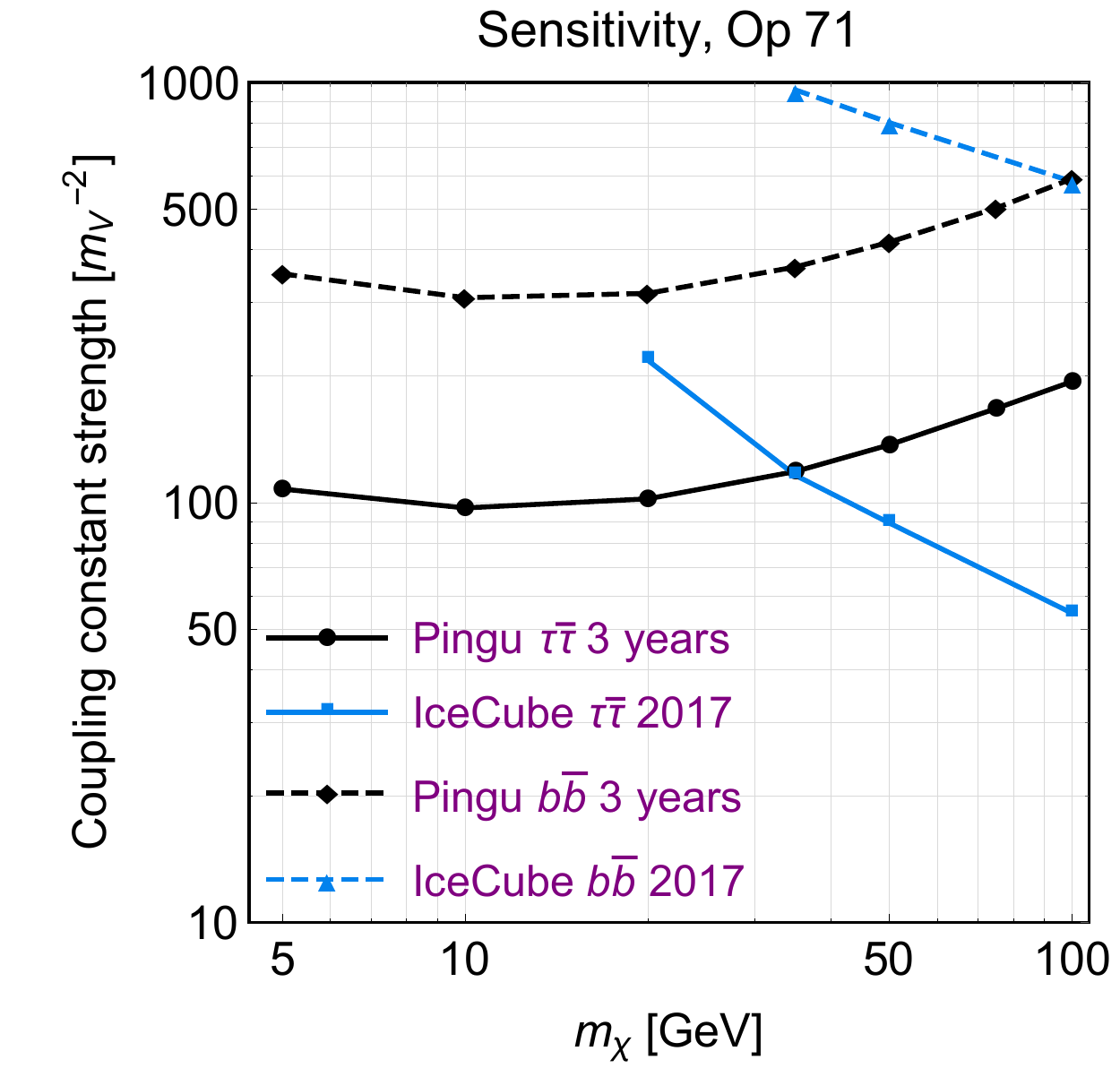}
\end{minipage}
\begin{minipage}[t]{0.32\linewidth}
\centering
\includegraphics[width=\textwidth]{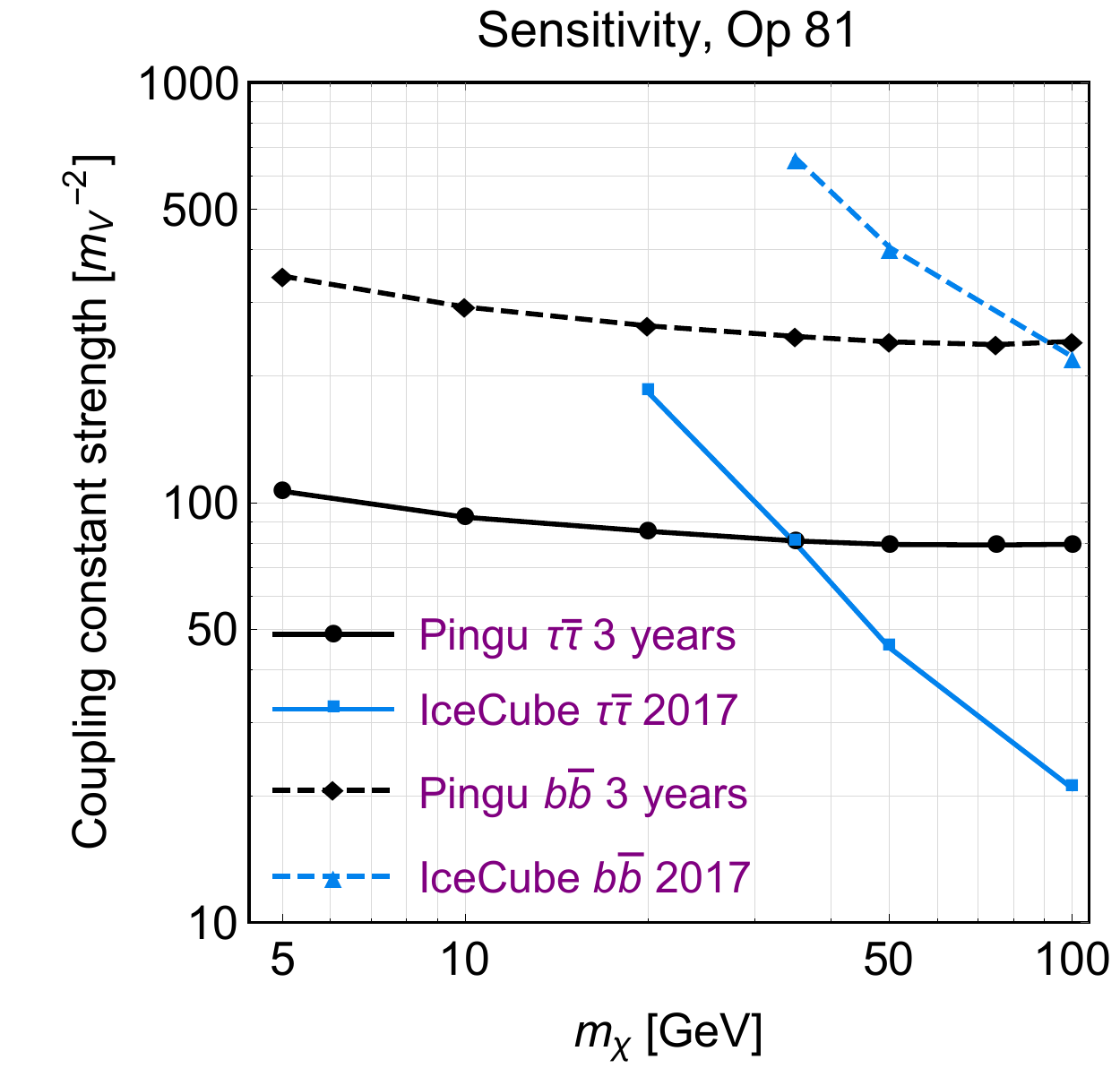}
\end{minipage}
\begin{minipage}[t]{0.32\linewidth}
\centering
\includegraphics[width=\textwidth]{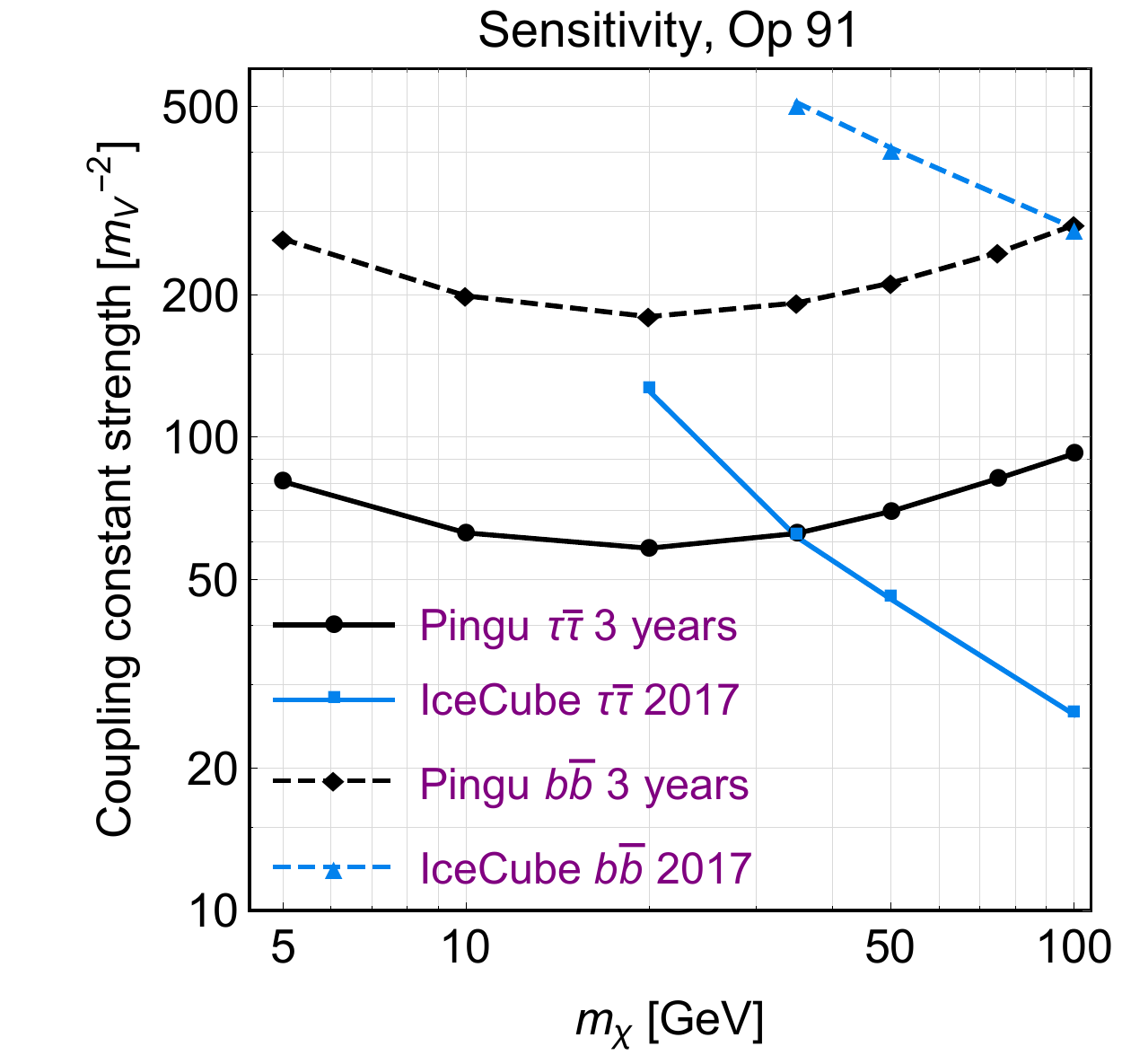}
\end{minipage}
\end{center}
\caption{Same as Fig.~\ref{fig:op1op4}, but now for the the isovector component of the operators $\hat{\mathcal{O}}_3=i{\bf{\hat{S}}}_N\cdot({\bf{\hat{q}}}/m_N)\times{\bf{\hat{v}}}^{\perp}\mathds{1}_\chi$, $\hat{\mathcal{O}}_5=i{\bf{\hat{S}}}_\chi\cdot[({{\bf{\hat{q}}}/m_N})\times{\bf{\hat{v}}}^{\perp}]\mathds{1}_N$, $\hat{\mathcal{O}}_6={\bf{\hat{S}}}_\chi\cdot({\bf{\hat{q}}}/m_N) {\bf{\hat{S}}}_N\cdot(\hat{{\bf{q}}}/m_N)$, $\hat{\mathcal{O}}_7={\bf{\hat{S}}}_{N}\cdot {\bf{\hat{v}}}^{\perp}\mathds{1}_\chi$, $\hat{\mathcal{O}}_8={\bf{\hat{S}}}_{\chi}\cdot {\bf{\hat{v}}}^{\perp}\mathds{1}_N$, and $\hat{\mathcal{O}}_9=i{\bf{\hat{S}}}_\chi\cdot[{\bf{\hat{S}}}_N\times({\bf{\hat{q}}}/m_N)]$.} 
\label{fig:isov1}
\end{figure}

\begin{figure}[t]
\begin{center}
\begin{minipage}[t]{0.32\linewidth}
\centering
\includegraphics[width=\textwidth]{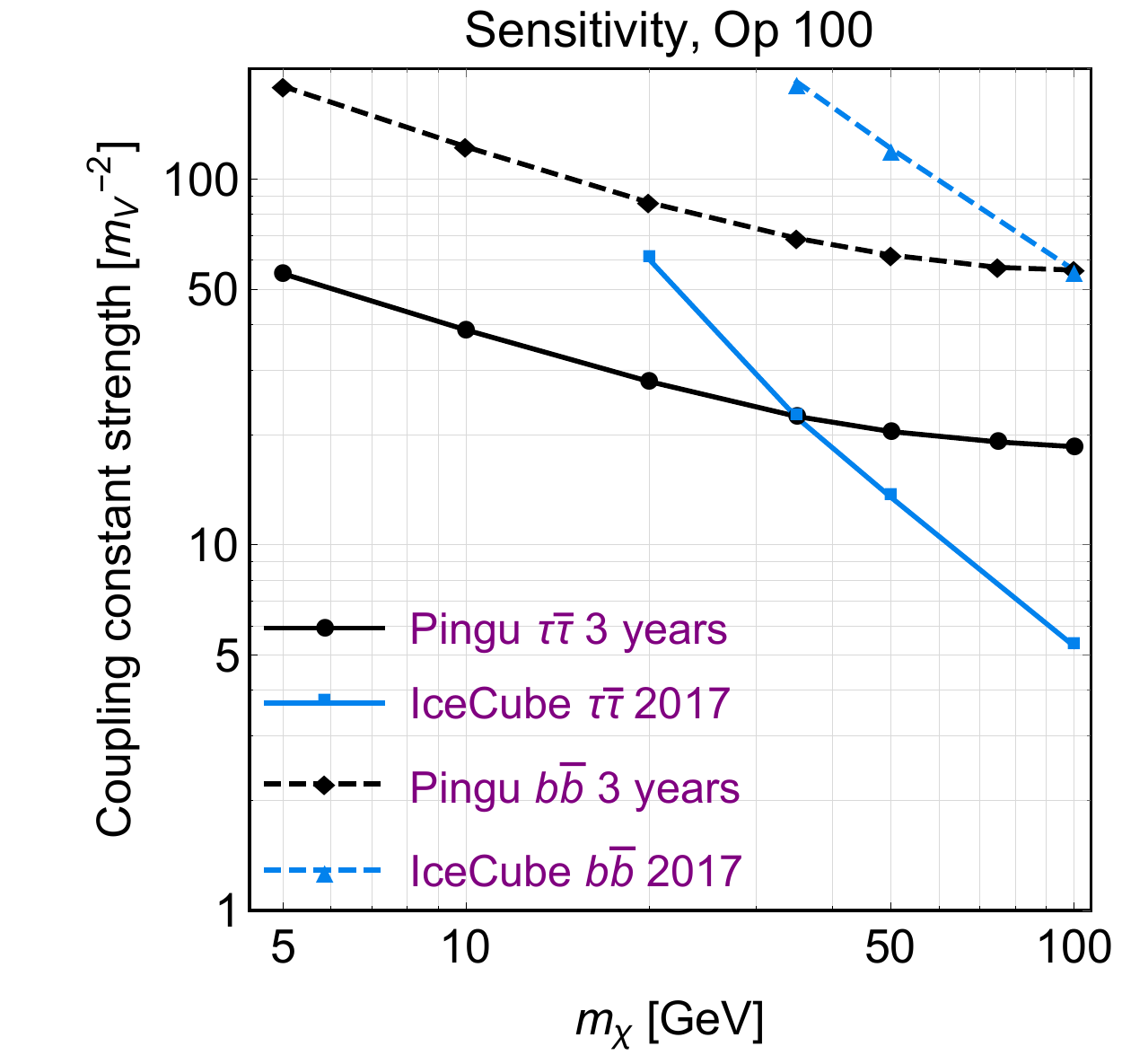}
\end{minipage}
\begin{minipage}[t]{0.32\linewidth}
\centering
\includegraphics[width=\textwidth]{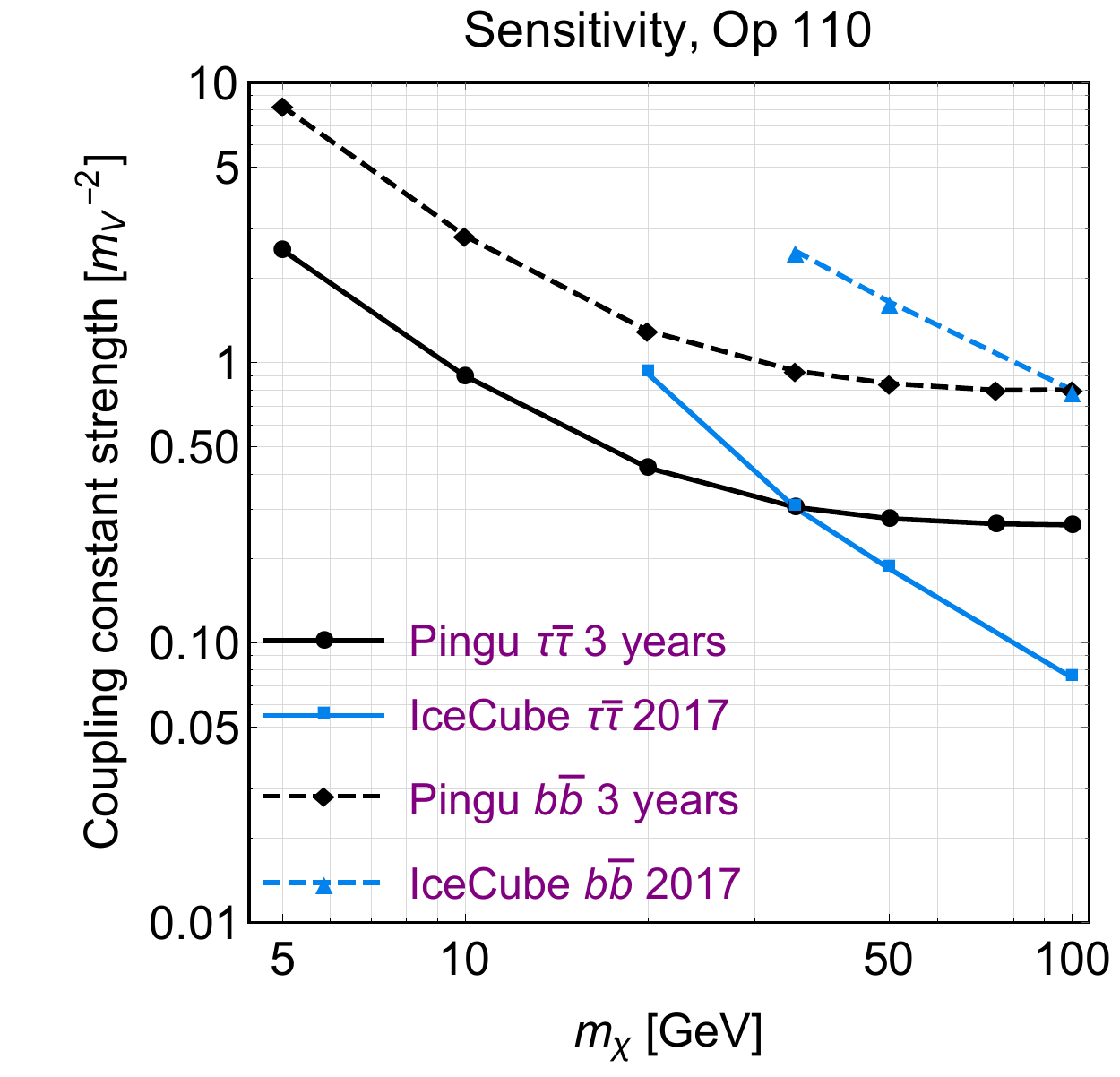}
\end{minipage}
\begin{minipage}[t]{0.32\linewidth}
\centering
\includegraphics[width=\textwidth]{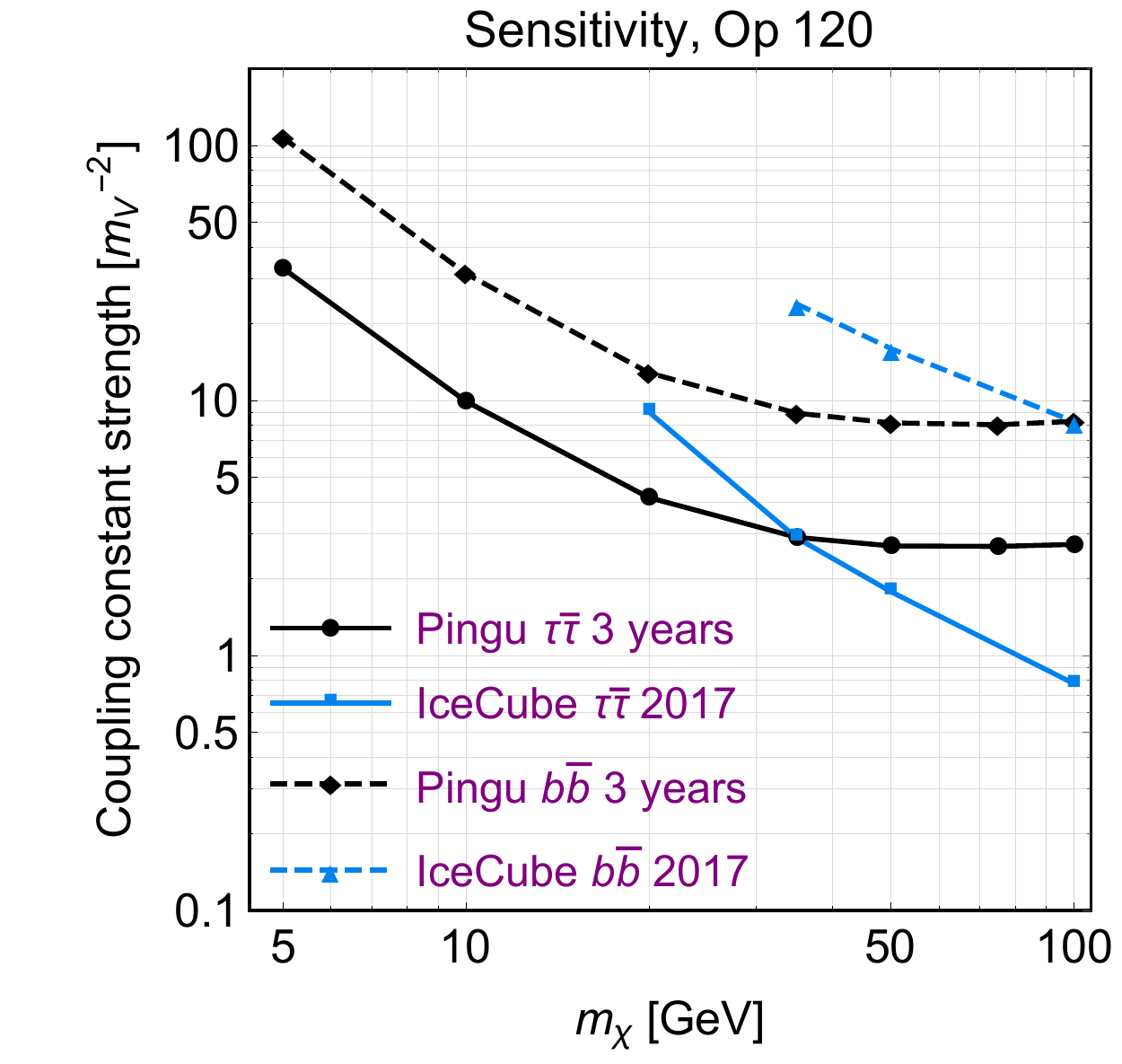}
\end{minipage}
\begin{minipage}[t]{0.32\linewidth}
\centering
\includegraphics[width=\textwidth]{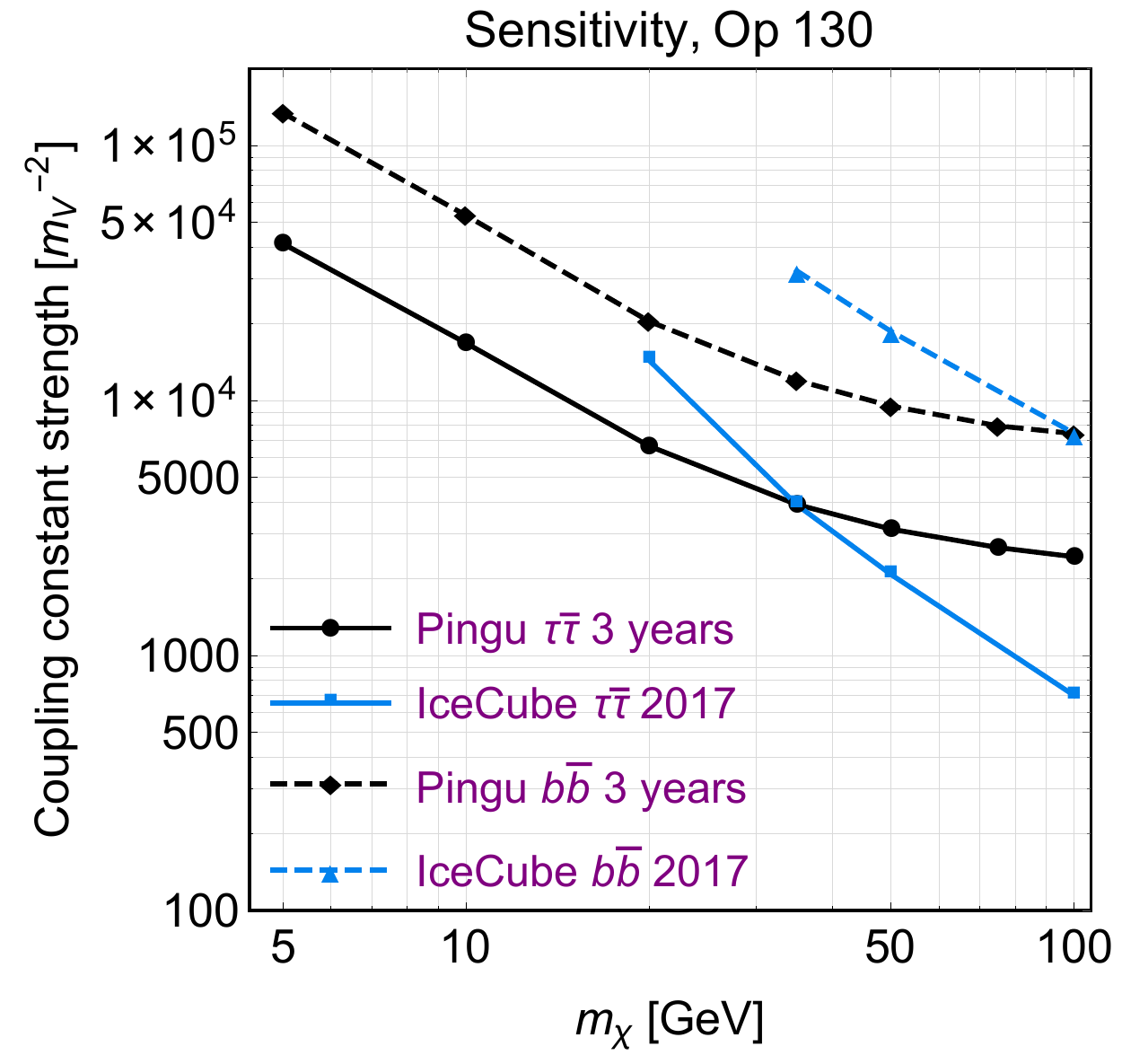}
\end{minipage}
\begin{minipage}[t]{0.32\linewidth}
\centering
\includegraphics[width=\textwidth]{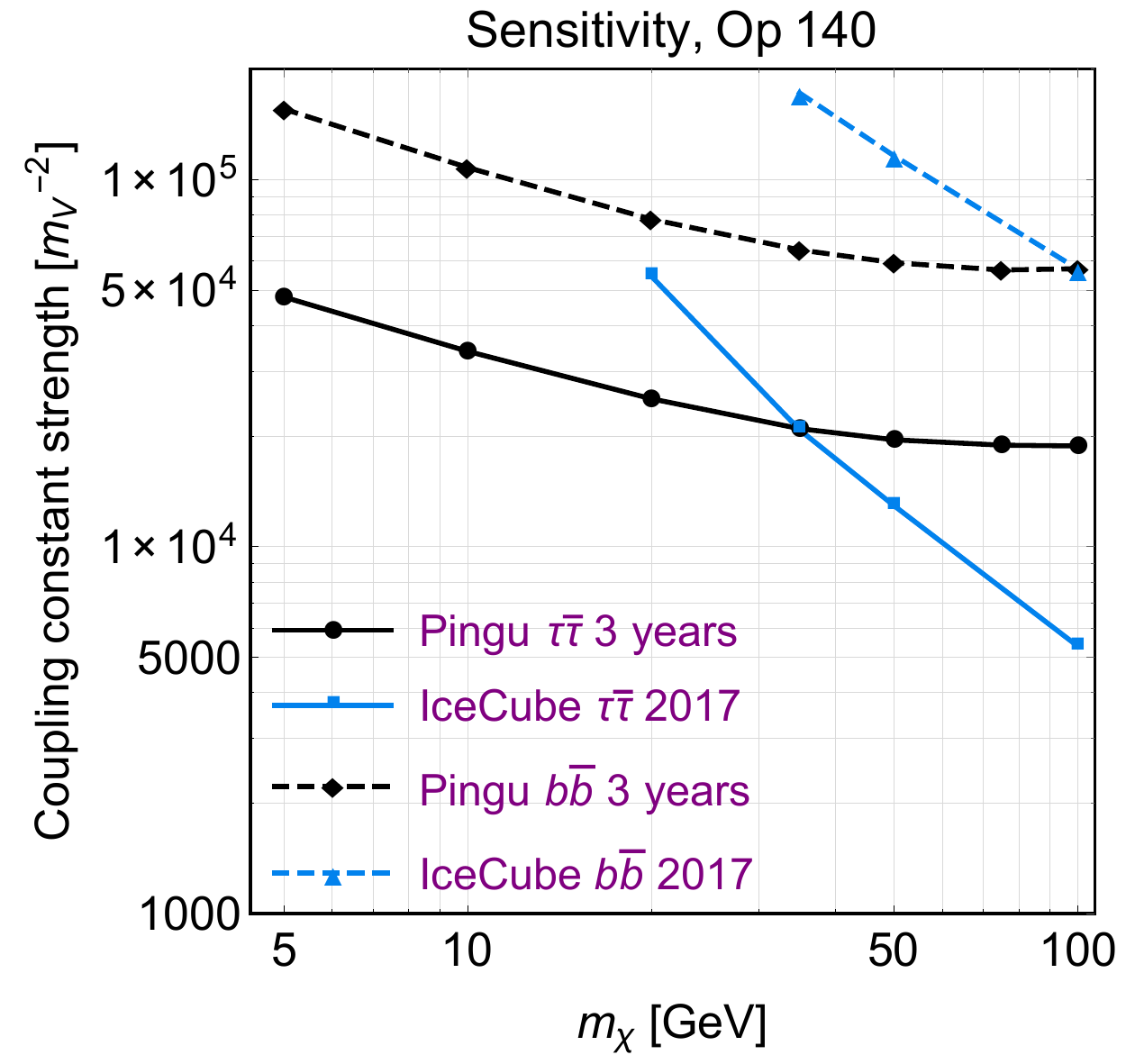}
\end{minipage}
\begin{minipage}[t]{0.32\linewidth}
\centering
\includegraphics[width=\textwidth]{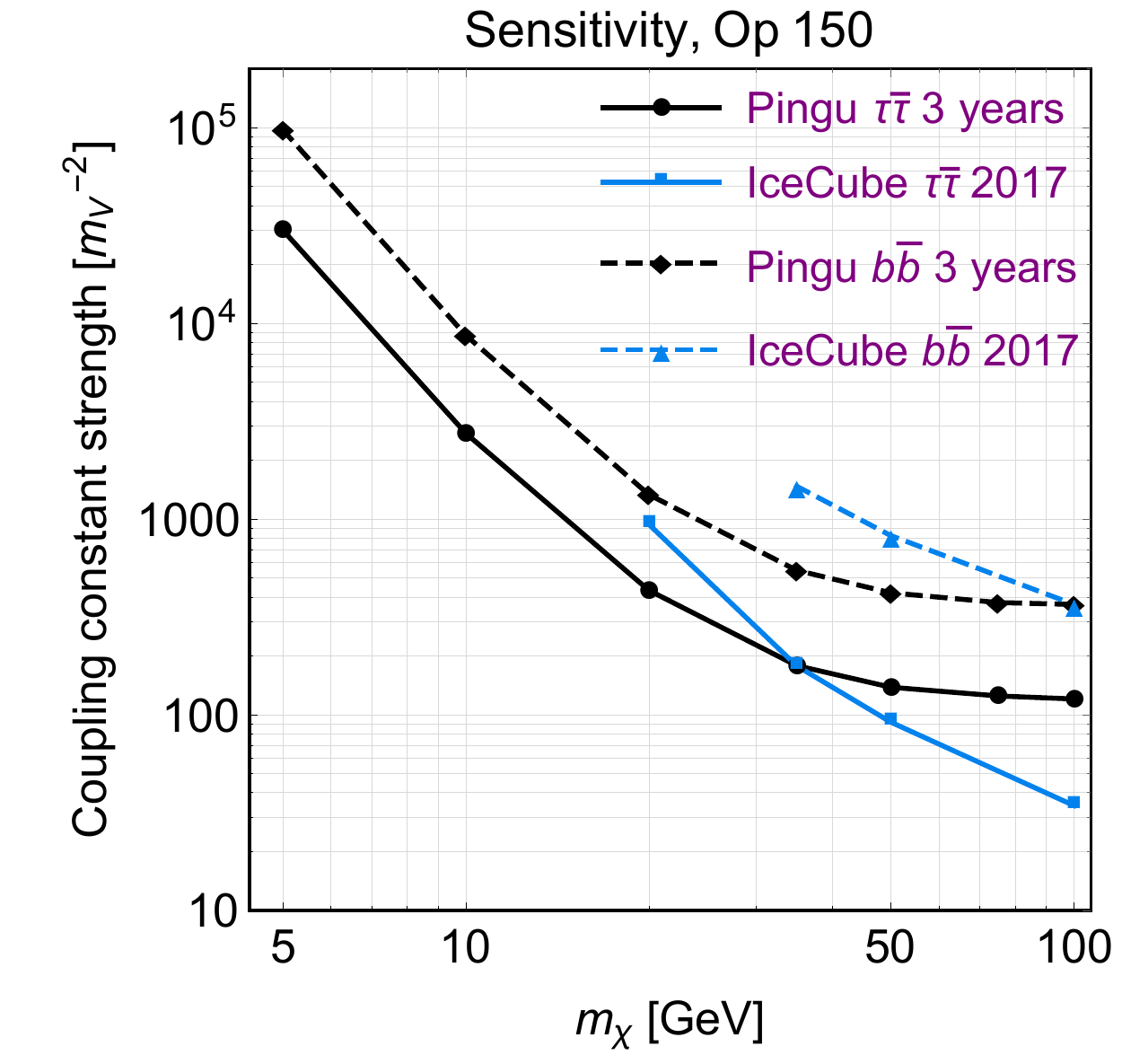}
\end{minipage}
\end{center}
\caption{Same as Fig.~\ref{fig:op1op4}, but now for the the isoscalar component of the operators $\hat{\mathcal{O}}_{10} = i{\bf{\hat{S}}}_N\cdot({\bf{\hat{q}}}/m_N)\mathds{1}_\chi$, $\hat{\mathcal{O}}_{11} = i{\bf{\hat{S}}}_\chi\cdot({\bf{\hat{q}}}/m_N)\mathds{1}_N$, $\hat{\mathcal{O}}_{12} = {\bf{\hat{S}}}_{\chi}\cdot({\bf{\hat{S}}}_{N} \times{\bf{\hat{v}}}^{\perp})$, $\hat{\mathcal{O}}_{13} =i ({\bf{\hat{S}}}_{\chi}\cdot {\bf{\hat{v}}}^{\perp})({\bf{\hat{S}}}_{N}\cdot {\bf{\hat{q}}}/m_N)$, $\hat{\mathcal{O}}_{14} = i{\bf{\hat{S}}}_{\chi}\cdot ({\bf{\hat{q}}}/m_N)({\bf{\hat{S}}}_{N}\cdot {\bf{\hat{v}}}^{\perp})$ and $\hat{\mathcal{O}}_{15} = -{\bf{\hat{S}}}_{\chi}\cdot ({\bf{\hat{q}}}/m_N)[ ({\bf{\hat{S}}}_{N}\times {\bf{\hat{v}}}^{\perp} ) \cdot ({\bf{\hat{q}}}/m_N)] $.}  
\label{fig:isos2}
\end{figure}

\begin{figure}[t]
\begin{center}
\begin{minipage}[t]{0.32\linewidth}
\centering
\includegraphics[width=\textwidth]{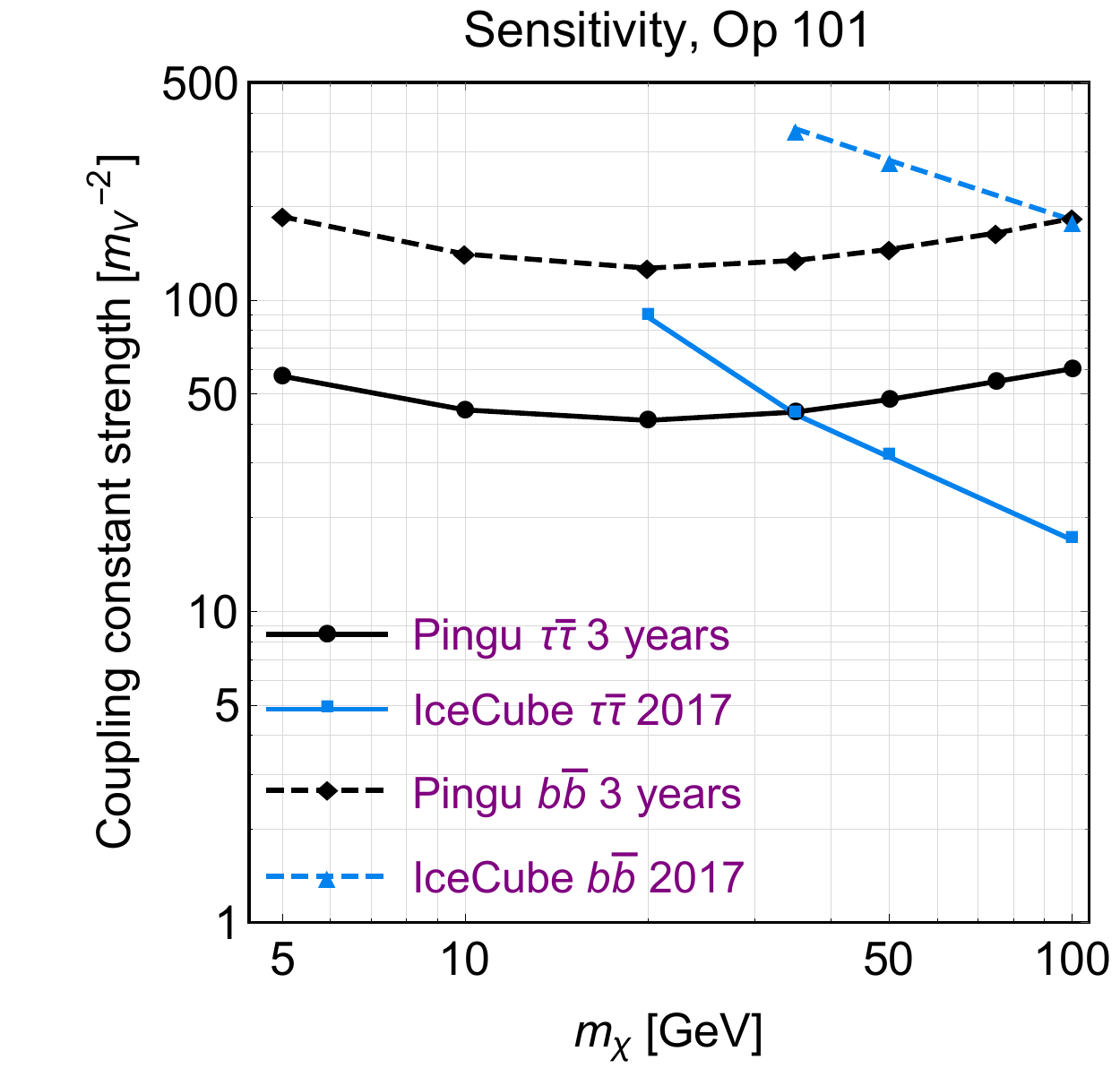}
\end{minipage}
\begin{minipage}[t]{0.32\linewidth}
\centering
\includegraphics[width=\textwidth]{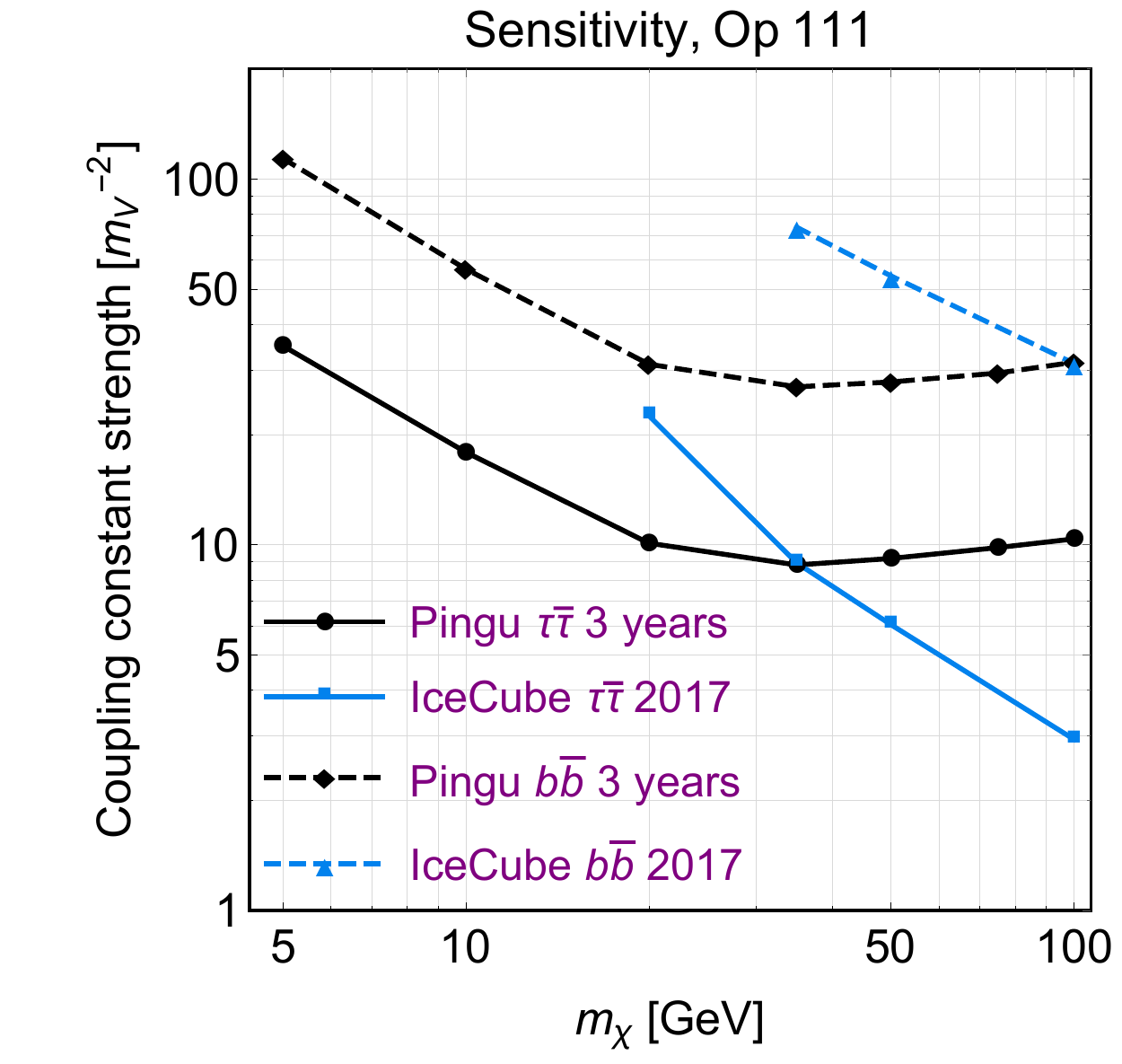}
\end{minipage}
\begin{minipage}[t]{0.32\linewidth}
\centering
\includegraphics[width=\textwidth]{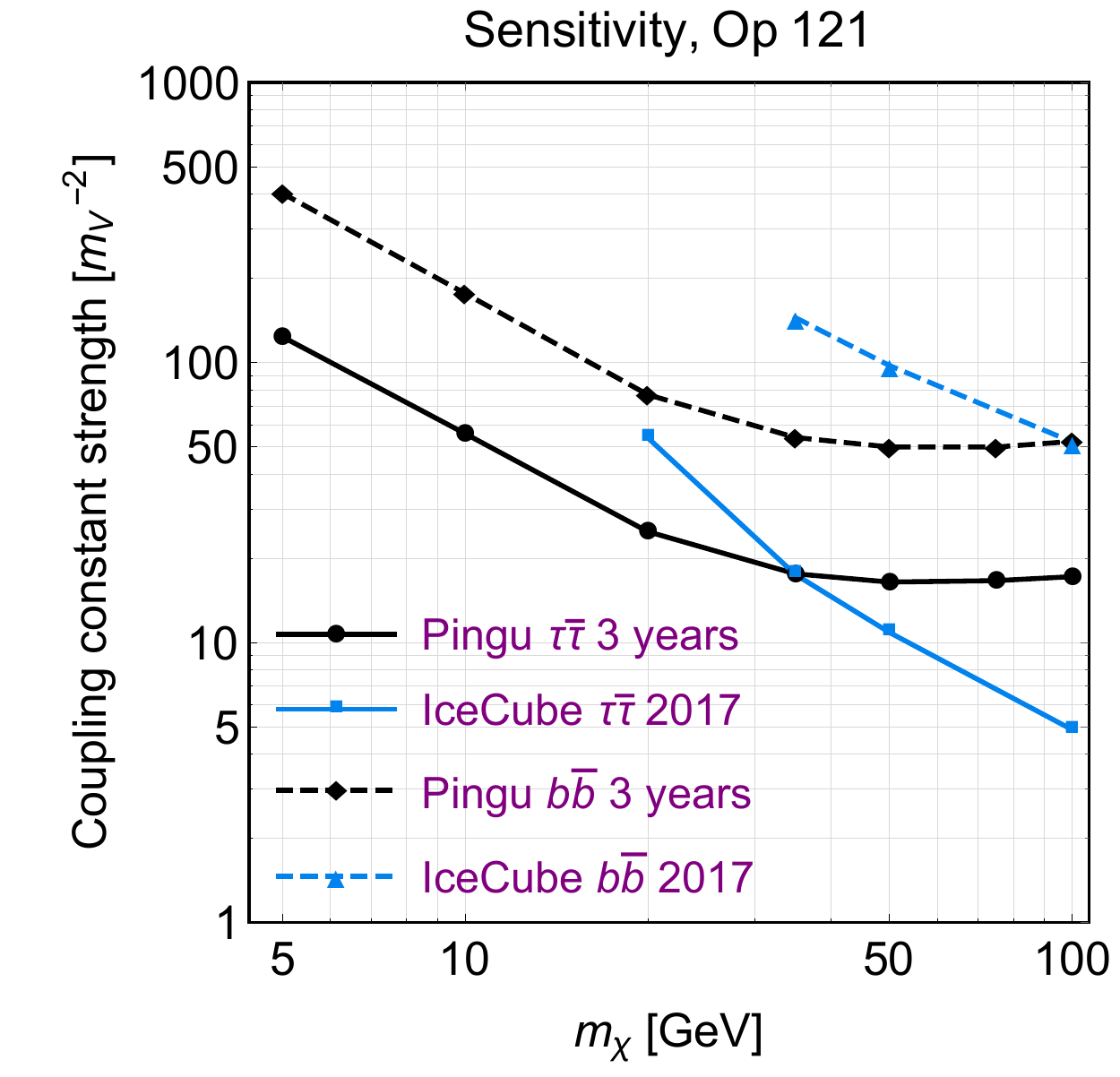}
\end{minipage}
\begin{minipage}[t]{0.32\linewidth}
\centering
\includegraphics[width=\textwidth]{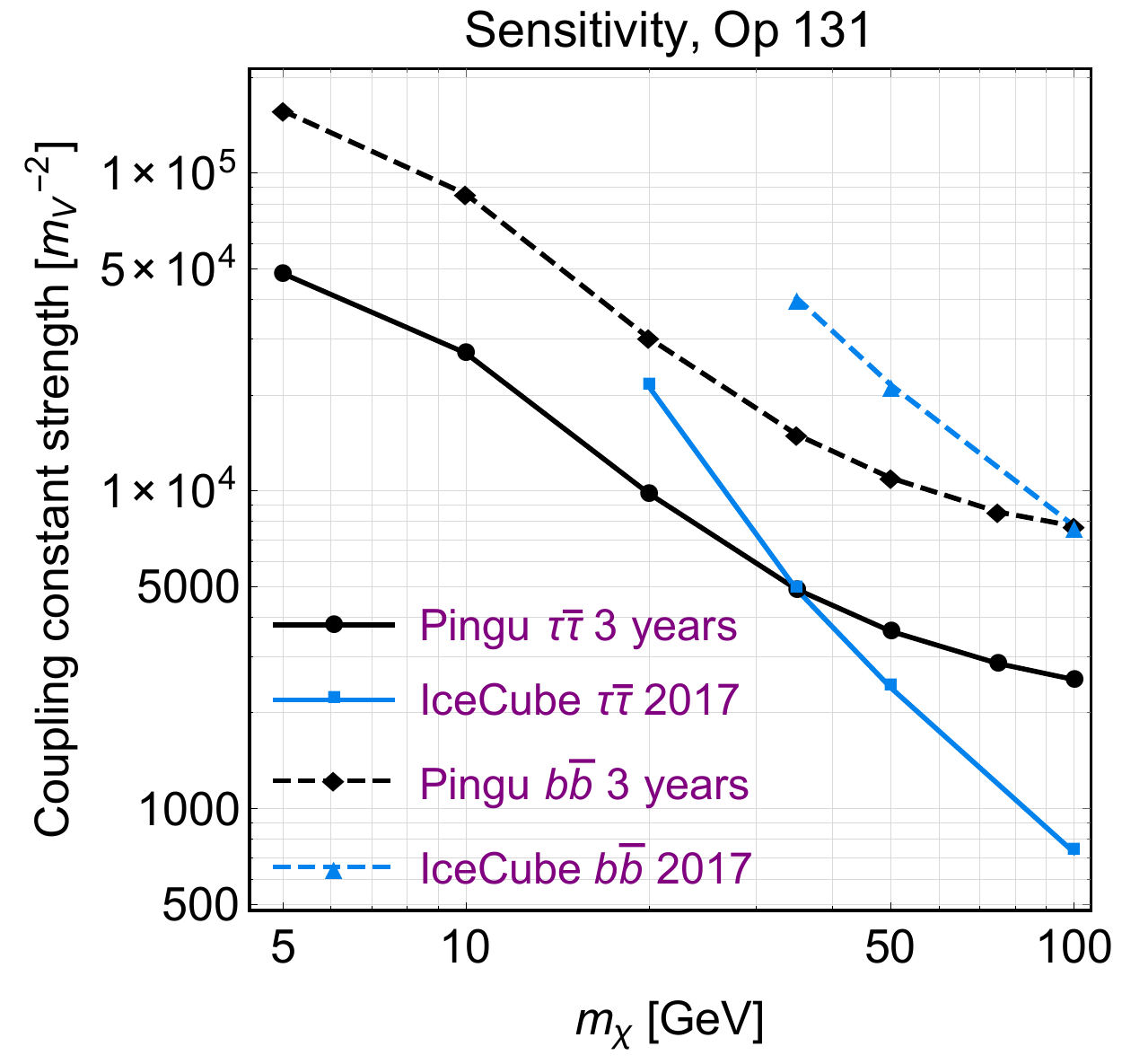}
\end{minipage}
\begin{minipage}[t]{0.32\linewidth}
\centering
\includegraphics[width=\textwidth]{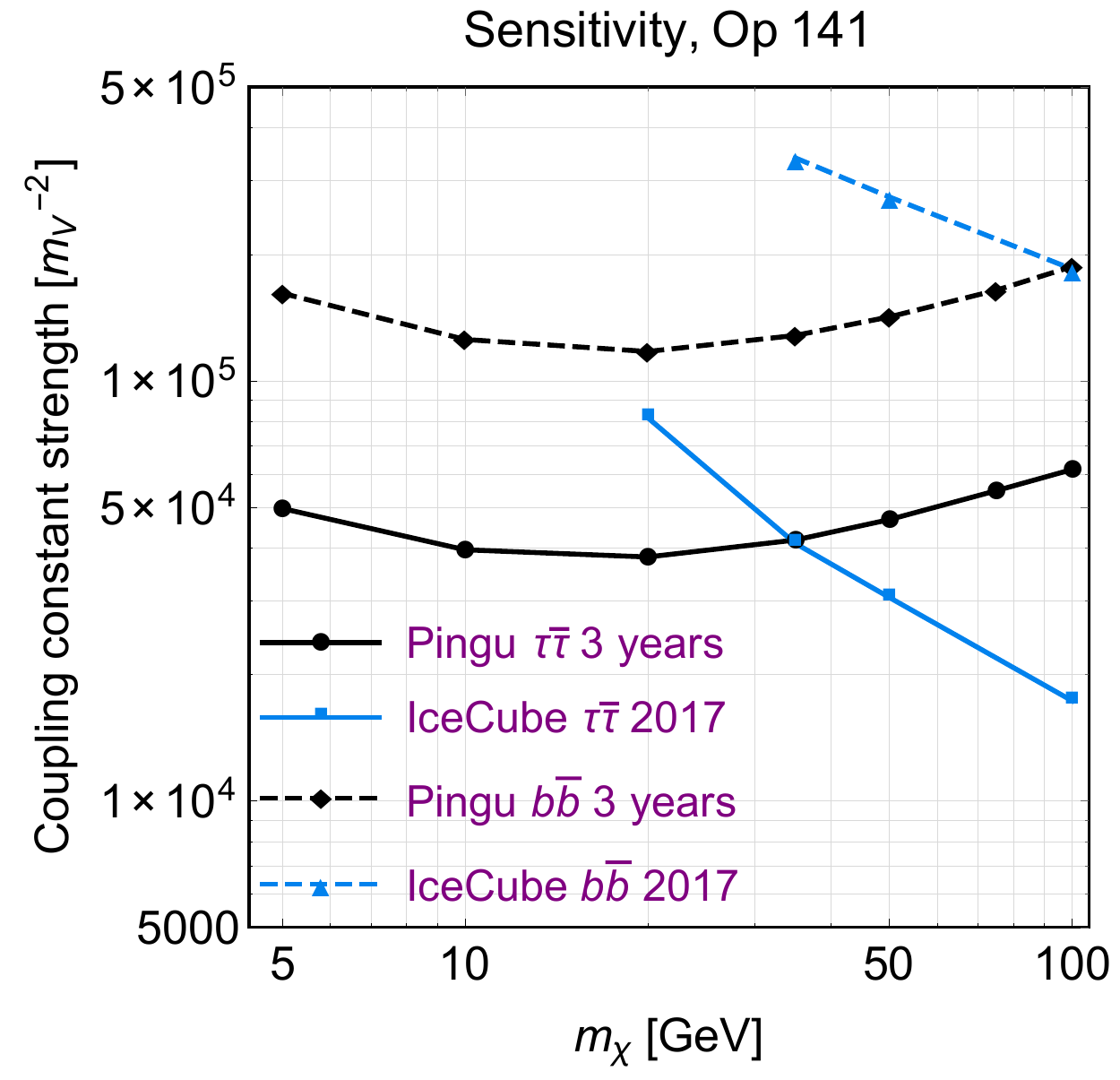}
\end{minipage}
\begin{minipage}[t]{0.32\linewidth}
\centering
\includegraphics[width=\textwidth]{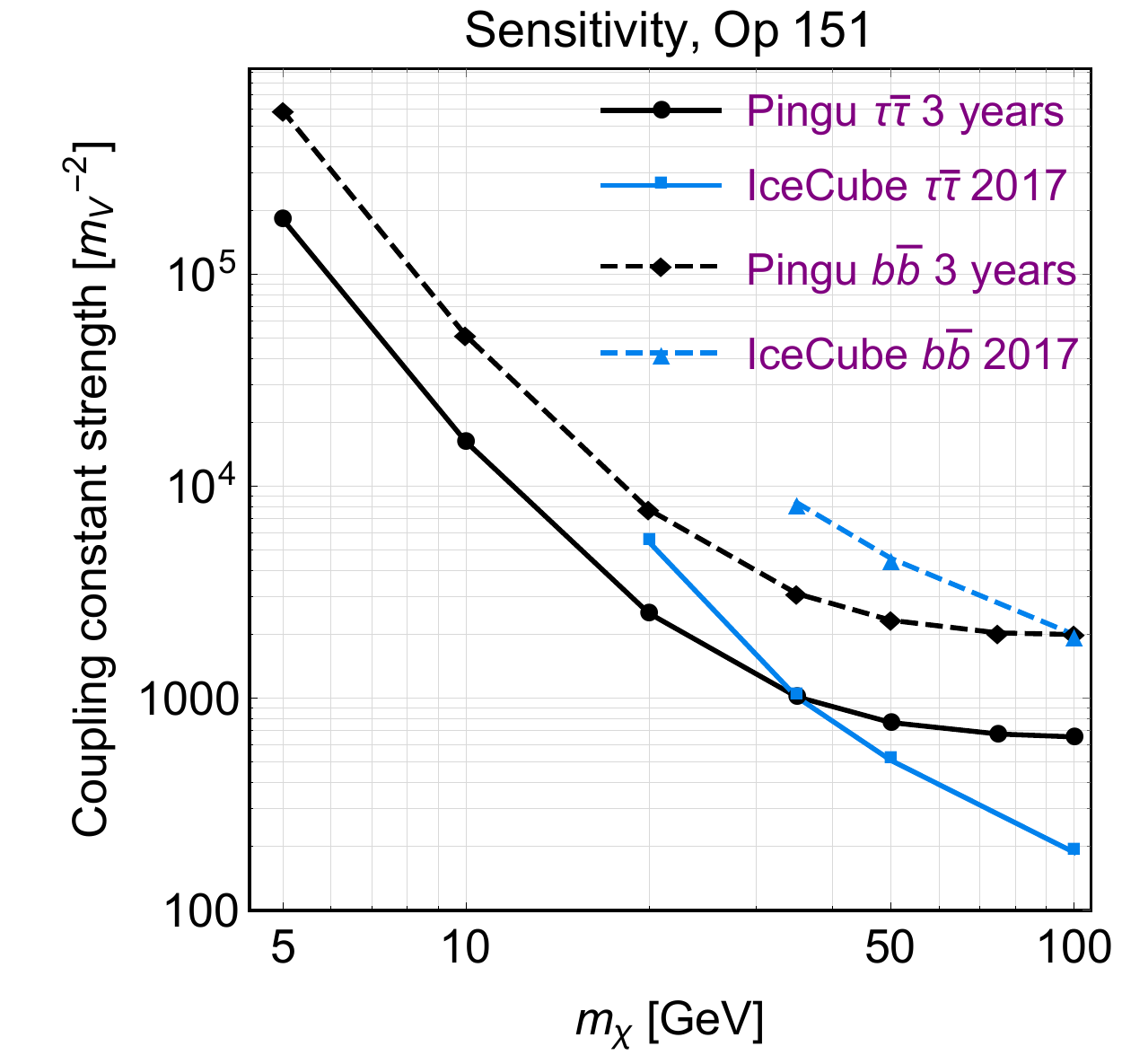}
\end{minipage}
\end{center}\caption{Same as Fig.~\ref{fig:op1op4}, but now for the the isovector component of the operators $\hat{\mathcal{O}}_{10} = i{\bf{\hat{S}}}_N\cdot({\bf{\hat{q}}}/m_N)\mathds{1}_\chi$, $\hat{\mathcal{O}}_{11} = i{\bf{\hat{S}}}_\chi\cdot({\bf{\hat{q}}}/m_N)\mathds{1}_N$, $\hat{\mathcal{O}}_{12} = {\bf{\hat{S}}}_{\chi}\cdot({\bf{\hat{S}}}_{N} \times{\bf{\hat{v}}}^{\perp})$, $\hat{\mathcal{O}}_{13} =i ({\bf{\hat{S}}}_{\chi}\cdot {\bf{\hat{v}}}^{\perp})({\bf{\hat{S}}}_{N}\cdot {\bf{\hat{q}}}/m_N)$, $\hat{\mathcal{O}}_{14} = i{\bf{\hat{S}}}_{\chi}\cdot ({\bf{\hat{q}}}/m_N)({\bf{\hat{S}}}_{N}\cdot {\bf{\hat{v}}}^{\perp})$ and $\hat{\mathcal{O}}_{15} = -{\bf{\hat{S}}}_{\chi}\cdot ({\bf{\hat{q}}}/m_N)[ ({\bf{\hat{S}}}_{N}\times {\bf{\hat{v}}}^{\perp} ) \cdot ({\bf{\hat{q}}}/m_N)] $.}  
\label{fig:isov2}
\end{figure}

\subsection{Effective area and angular resolution}
\label{sec:exp}
A first experimental input that we need to evaluate for Eqs.~(\ref{eq:signal}) and (\ref{eq:bg}) is the effective area.~The effective area used in  Eqs.~(\ref{eq:signal}) and (\ref{eq:bg}) depends on the assumed detector and is defined as the area for which the detector observes a neutrino with 100\% efficiency.~In this analysis, we assume the effective areas given in~\cite{IC3} for IceCube and DeepCore, and the effective mass reported in~\cite{LoI} for PINGU.~Effective mass and area are related by the following expression
\begin{equation} 
A^{\mathrm{eff}}_\nu=\dfrac{m_{\mathrm{eff}}}{M_{\mathrm{ice}}}\cdot N_a(n_p\sigma_{\nu p}+n_n\sigma_{\nu n})\,, \qquad\qquad \nu=\nu_\mu,\bar{\nu}_{\mu} 
\label{eq:effpingu}
\end{equation}
where $m_{\mathrm{eff}}$ is the effective mass (defined as ice density times effective detector volume), $M_\mathrm{ice}$ is the mass of ice per mole, $N_a$ is Avogadro's number, $n_n$ ($n_p$) is the number of neutrons (protons) in a ice molecule, and $\sigma_{\nu n}$ ($\sigma_{\nu p}$) is the cross section for neutrino-neutron (-proton) scattering.~These cross sections were taken from \cite{Gandhi:1995tf}.~Fig.~\ref{EffArea} shows the effective areas used in this analysis.~To obtain Fig.~\ref{EffArea}, we extrapolated PINGU's effective mass~\cite{LoI} from 50 GeV to 100 GeV.

The second experimental input that we need is the median angular resolution of the assumed detector, $\theta_{\rm res}$ in Eqs.~(\ref{eq:signal}) and (\ref{eq:bg}).~Here we use the angular resolution given in \cite{IC3} for the DeepCore and IceCube detectors, and the one reported in~\cite{LoI} for PINGU.~Since the angular resolution is only given up to 29 GeV in~\cite{LoI}, we assume that it remains flat from 29 to 100 GeV.

\section{Results}
\label{sec:results}
In this section, we calculate PINGU's sensitivity to the 28 coupling constants that characterise the non-relativistic effective theory of (spin 1/2) DM-nucleon interactions introduced in Sec.~\ref{sec:theory} (and references therein).~We express PINGU's sensitivity in terms of 5$\sigma$ contours in the DM-mass -- coupling constant plane.~These correspond to a significance of $S=5$, where $S$ is given in Eq.~(\ref{eq:sign1}).~For each DM-nucleon interaction type, we compute the sensitivity contours by solving the equation $S=5$ for the corresponding coupling constant at each value of the DM particle mass in the 5 - 100 GeV range, and assuming three years of continuous data taking.~For each DM mass -- coupling constant pair, we compare our sensitivity contours with 90\% C.L. exclusion limits that we obtain from a reanalysis of the null result of current DM searches at IceCube/DeepCore~\cite{IC3}.~For each DM-nucleon interaction type, we separately consider two cases.~In the first one, DM annihilates into $b\bar{b}$ pairs, producing a soft neutrino energy spectrum.~In the second one, DM annihilates into $\tau\bar{\tau}$ pairs, generating a hard neutrino energy spectrum.

Fig.~\ref{fig:op1op4} shows PINGU's 5$\sigma$ sensitivity contours together with 90\% C.L. exclusion limits from the null result of DM searches at IceCube/DeepCore for the interaction operators $\hat{\mathcal{O}}_1=\mathds{1}_{\chi}\mathds{1}_N$ and $\hat{\mathcal{O}}_4={\bf{\hat{S}}}_{\chi}\cdot {\bf{\hat{S}}}_{N}$.~These correspond to canonical spin-independent and spin-dependent interactions, respectively.~The left panels refer to the isoscalar component of the two interaction operators, whereas the right panels correspond to the isovector counterpart.~On the $y$-axis, coupling constants are expressed in units of the electroweak scale, here denoted by $m_V=246.2$~GeV.~For comparison, $c_1^0=10^{-3}/m_V^2$ corresponds to a spin-independent cross section for DM-nucleon scattering equal to $10^{-6}\mu_{\chi N}^2/(4\pi m_V^4)$, where $\mu_{\chi N}$ is the DM-nucleon reduced mass.~Evaluating this expression at $m_\chi=50$~GeV, one finds the value $8\times10^{-45}$ cm$^{-2}$.

Assuming that DM primarily annihilates into $\tau\bar{\tau}$ pairs (hard neutrino energy spectrum), we find that PINGU will be able to detect values of $c_1^0$ as low as $10^{-2}/m_V^2$ with a statistical significance of at least $5\sigma$ for $m_{\chi} \sim 50$~GeV.~Similarly, it will be able to detect values of $c_4^0$ as low as $3\times 10^{-2}/m_V^2$ with a statistical significance of at least $5\sigma$ for $m_{\chi} \sim 10$~GeV.~When DM primarily annihilates into $b\bar{b}$ pairs (soft neutrino energy spectrum), the smallest coupling constant that PINGU can detect with a statistical significance of $5\sigma$ is a factor of $\sim 3$ larger, both for $c_1^0$ and  for $c_4^0$.

While PINGU's sensitivity contours and IceCube/DeepCore 90\% C.L. exclusion limits intersects at $m_\chi \sim 35$~GeV in the case of DM annihilation into $\tau\bar{\tau}$ (both for $\hat{\mathcal{O}}_1$ and for $\hat{\mathcal{O}}_4$), sensitivity contours extend far below the corresponding exclusion limits in the case of DM annihilation into $b\bar{b}$.~This shows that by lowering the experimental energy threshold from $E_{\rm th}\sim10$~GeV, as in the case of IceCube/DeepCore (see Fig.~\ref{EffArea}), to $E_{\rm th}\sim 1$~GeV, as in the case of PINGU, leads to a dramatic improvement in terms of sensitivity to DM-nucleon interactions, especially when the expected neutrino energy spectrum is soft, as for DM annihilation into $b\bar{b}$.

Let us now compare the results that we obtain for isoscalar interactions with those found for isovector couplings.~Fig.~\ref{fig:op1op4} shows that PINGU's sensitivity to isoscalar couplings is greater than to isovector couplings, and that IceCube/DeepCore 90\% C.L. exclusion limits on the former are stronger than on the latter, but not for the interaction operators that couple DM to the nuclear spin current, like $\hat{\mathcal{O}}_4$.~This result is expected since in the case of isovector couplings, the constants for DM coupling to protons and neutrons have opposite signs, which implies a cancellation in the coherently enhanced nuclear matrix elements upon which scattering amplitudes depend.~This results in suppressed DM-nucleus scattering cross sections, lower sensitivity and weaker limits for $\hat{\mathcal{O}}_1$.~In contrast, for operators that couple DM to the nuclear spin current, the above cancellation does not occur, even in the case of isovector couplings~\cite{Catena:2015iea}.

Let us now focus on DM-nucleon interaction operators different from the canonical $\hat{\mathcal{O}}_{1}$ and $\hat{\mathcal{O}}_{4}$.~Fig.~\ref{fig:isos1} shows PINGU's 5$\sigma$ sensitivity contours together with 90\% C.L. exclusion limits from the null result of DM searches at IceCube/DeepCore for the isoscalar component of the interaction operators $\hat{\mathcal{O}}_3=i{\bf{\hat{S}}}_N\cdot({\bf{\hat{q}}}/m_N)\times{\bf{\hat{v}}}^{\perp}\mathds{1}_\chi$, $\hat{\mathcal{O}}_5=i{\bf{\hat{S}}}_\chi\cdot[({{\bf{\hat{q}}}/m_N})\times{\bf{\hat{v}}}^{\perp}]\mathds{1}_N$, $\hat{\mathcal{O}}_6={\bf{\hat{S}}}_\chi\cdot({\bf{\hat{q}}}/m_N) {\bf{\hat{S}}}_N\cdot(\hat{{\bf{q}}}/m_N)$, $\hat{\mathcal{O}}_7={\bf{\hat{S}}}_{N}\cdot {\bf{\hat{v}}}^{\perp}\mathds{1}_\chi$, $\hat{\mathcal{O}}_8={\bf{\hat{S}}}_{\chi}\cdot {\bf{\hat{v}}}^{\perp}\mathds{1}_N$, and $\hat{\mathcal{O}}_9=i{\bf{\hat{S}}}_\chi\cdot[{\bf{\hat{S}}}_N\times({\bf{\hat{q}}}/m_N)]$ (see Tab.~\ref{tab:operators}).~Among these operators, $\hat{\mathcal{O}}_7$ and $\hat{\mathcal{O}}_8$ are the only ones generating a differential cross section for DM-nucleus scattering which does not go to zero in the limit of zero momentum transfer (see Eq.~(\ref{eq:R})).~Within this set of interaction operators, PINGU is particularly sensitive to $\hat{\mathcal{O}}_8$ (and $\hat{\mathcal{O}}_3$, but only for $m_\chi\sim50$~GeV).~PINGU has a good sensitivity to $\hat{\mathcal{O}}_8$ since this operator generates a coherently enhanced DM-nucleus scattering cross section which in the zero momentum transfer limit is proportional to the squared of the number of nucleons bound in the nucleus (similarly to the $\hat{\mathcal{O}}_1$ operator).~Fig.~\ref{fig:isov1} shows PINGU's 5$\sigma$ sensitivity contours together with 90\% C.L. exclusion limits from the null result of DM searches at IceCube/DeepCore for the isovector component of the above interaction operators.~Similarly to the case of the interaction operators $\hat{\mathcal{O}}_1$ and $\hat{\mathcal{O}}_4$ (see Fig.~\ref{fig:op1op4}), the sensitivity of PINGU to isovector DM-nucleon interactions is lower, as compared to PINGU's sensitivity to the corresponding isoscalar interactions, but not for the interaction operators that couple DM to the nuclear spin current, like $\hat{\mathcal{O}}_7$ and $\hat{\mathcal{O}}_9$.~For these operators, cancellations between proton and neutron contributions to the scattering cross section do not occur, even in the case of isovector couplings~\cite{Catena:2015iea}.

Finally, Fig.~\ref{fig:isos2} and Fig.~\ref{fig:isov2} show PINGU's 5$\sigma$ sensitivity contours together with 90\% C.L. exclusion limits from the null result of DM searches at IceCube/DeepCore for, respectively, the isoscalar and isovector component of the interaction operators $\hat{\mathcal{O}}_{10} = i{\bf{\hat{S}}}_N\cdot({\bf{\hat{q}}}/m_N)\mathds{1}_\chi$, $\hat{\mathcal{O}}_{11} = i{\bf{\hat{S}}}_\chi\cdot({\bf{\hat{q}}}/m_N)\mathds{1}_N$, $\hat{\mathcal{O}}_{12} = {\bf{\hat{S}}}_{\chi}\cdot({\bf{\hat{S}}}_{N} \times{\bf{\hat{v}}}^{\perp})$, $\hat{\mathcal{O}}_{13} =i ({\bf{\hat{S}}}_{\chi}\cdot {\bf{\hat{v}}}^{\perp})({\bf{\hat{S}}}_{N}\cdot {\bf{\hat{q}}}/m_N)$, $\hat{\mathcal{O}}_{14} = i{\bf{\hat{S}}}_{\chi}\cdot ({\bf{\hat{q}}}/m_N)({\bf{\hat{S}}}_{N}\cdot {\bf{\hat{v}}}^{\perp})$ and $\hat{\mathcal{O}}_{15} = -{\bf{\hat{S}}}_{\chi}\cdot ({\bf{\hat{q}}}/m_N)[ ({\bf{\hat{S}}}_{N}\times {\bf{\hat{v}}}^{\perp} ) \cdot ({\bf{\hat{q}}}/m_N)] $.~Within this set of interaction operators, PINGU is particularly sensitive to the operator $\hat{\mathcal{O}}_{11}$, which, analogously to the $\hat{\mathcal{O}}_1$ operator, generates a coherent DM-nucleus scattering with cross section scaling like $A^2$ in the zero momentum transfer limit (where $A$ is the number of nucleons bound in the nucleus).

We refer to~\cite{Kang:2018odb} for a comparison between our results and current exclusion limits from direct detection experiments.~As already noticed in~\cite{Catena:2015iea}, compared to direct detection experiments neutrino telescopes can currently probe smaller values of the DM-nucleon coupling constants for some of the spin-dependent interactions in Tab.~\ref{tab:operators}, and for the velocity-dependent operator $\hat{\mathcal{O}}_7$.

\section{Conclusion}
\label{sec:conclusion}
In this work, we calculated PINGU's sensitivity to the 28 coupling constants characterising the non-relativistic effective theory of (spin 1/2) DM-nucleon interactions introduced in Sec.~\ref{sec:theory}.~In this calculation, we expressed PINGU's sensitivity in terms of $5\sigma$ sensitivity contours in the DM-mass -- coupling constant plane, focusing on DM masses in the 5 - 100 GeV range.~For each DM mass -- coupling constant pair, we then compared our sensitivity contours with the 90\% C.L. exclusion limits on the same (isoscalar and isovector) coupling constants that we obtained from a reanalysis of the null result of current searches for DM at IceCube/DeepCore.~For PINGU we assumed three years of continuous data taking, while for IceCube/DeepCore we considered 532 days of data taking (discarding periods of the year when the Sun is above horizon).~For each DM-nucleon interaction type, we considered two cases separately:~in the first one, DM primarily annihilates into $b\bar{b}$ pairs; in the second one, DM primarily annihilates into $\tau\bar{\tau}$ pairs.

We found that PINGU will be able to detect values of $c_1^0$ (i.e.~canonical isoscalar spin-independent interactions) as low as $10^{-2}/m_V^2$ with a statistical significance of at least $5\sigma$ for $m_{\chi} \sim 50$~GeV, if DM primarily annihilates into $\tau\bar{\tau}$ pairs (producing a hard neutrino energy spectrum).~Similarly, if DM primarily annihilates into $\tau\bar{\tau}$ pairs, PINGU will be able to detect values of $c_4^0$ (i.e.~canonical isoscalar spin-dependent interactions) as low as $3\times 10^{-2}/m_V^2$ with a statistical significance of at least $5\sigma$ for $m_{\chi}$ around 10~GeV.~The smallest coupling constant that PINGU can detect with a statistical significance of $5\sigma$ is a factor of $\sim 3$ larger, both for $c_1^0$ and  for $c_4^0$, when DM primarily annihilates into $b\bar{b}$ pairs (producing a soft neutrino energy spectrum).

We then extended our analysis of the canonical spin-independent and spin-dependent interactions to all DM-nucleon interaction operators in the effective theory of (spin 1/2) DM-nucleon interactions.~Within this large set of interaction operators, we found that PINGU is particularly sensitive to the interaction operators $\hat{\mathcal{O}}_8$ and $\hat{\mathcal{O}}_{11}$, which, analogously to the $\hat{\mathcal{O}}_1$ operator, generate a coherently enhanced DM-nucleus scattering cross section.~For each non-relativistic interaction operator, we presented our results for both isoscalar and isovector couplings.

Comparing isoscalar and isovector interactions, we found that for isovector couplings, PINGU sensitivity is lower than for isoscalar couplings when the DM-nucleus interaction is coherently enhanced, but it is not suppressed for interactions that couple DM to the nuclear spin current.~This result is expected, and can be explained as follows.~While in the isoscalar case, the constants for DM coupling to protons and neutrons have the same sign, in the isovector case the above coupling constants have opposite signs.~This implies a cancellation in the coherently enhanced nuclear matrix elements upon which scattering amplitudes depend, and, therefore, a lower sensitivity to the DM-nucleon interactions studied here.~On the other end, such a cancellation does not occur for interactions coupling DM to the nuclear spin current.

Finally, we found that PINGU's sensitivity contours are significantly below current IceCube/DeepCore 90\% C.L. exclusion limits when $b\bar{b}$ is the leading DM annihilation channel.~This conclusion holds true independently of the assumed DM-nucleon interaction.~It thus shows that lowering the experimental energy threshold from $E_{\rm th}\sim10$~GeV, to $E_{\rm th}\sim 1$~GeV, as in the case of PINGU, will improve the current sensitivity of neutrino telescopes to any DM-nucleon interaction dramatically, when $b\bar{b}$ is the leading DM annihilation channel.~When DM primarily annihilates into $\tau\bar{\tau}$ pair, the value of the DM particle mass below which PINGU will improve upon current exclusion limits is $m_\chi\sim 35$~GeV, independently of the assumed DM-nucleon interaction.

\acknowledgments A.B. and R.C would like to thank Anastasia Danopoulou, Fredrik Hellstr\"om, K\aa re Fridell, Martin B.~Krauss and Vanessa Zema for interesting discussions on dark matter capture in the Sun and related topics.~This work was supported by the Knut and Alice Wallenberg Foundation and is partly performed within the Swedish Consortium for Dark Matter Direct Detection (SweDCube).


\providecommand{\href}[2]{#2}\begingroup\raggedright\endgroup

\end{document}